\documentclass[hyper,a4paper]{JHEP3}

\usepackage{url}
\usepackage{epsfig}
\usepackage{amsmath}
\usepackage{amssymb}
\usepackage{graphics,bm}
\newcommand{\beq}{\begin{equation}}
\newcommand{\eeq}{\end{equation}}
\newcommand{\bqa}{\begin{eqnarray}}
\newcommand{\eqa}{\end{eqnarray}}
\def\sumint{\hbox{$\sum$}\!\!\!\!\!\!\int}

\title{Pressure to order $g^8\log{g}$ of massless $\phi^4$ theory at weak coupling}

\author{Jens O. Andersen\\Niels Bohr International Academy, Niels Bohr Institute, Blegdamsvej 17, DK-2100 Copenhagen, Denmark\\ On leave from: Department of Physics, Norwegian University of Science and Technology, H\o gskoleringen 5, N-7491 Trondheim, Norway\\ E-mail: \email{andersen@tf.phys.ntnu.no}}
\author{Lars T. Kyllingstad\\ Department of Physics, Norwegian University of Science and Technology, H\o gskoleringen 5, N-7491 Trondheim, Norway\\ E-mail: \email{lars.kyllingstad@ntnu.no}}
\author{Lars E. Leganger\\ Department of Physics, Norwegian University of Science and Technology, H\o gskoleringen 5, N-7491 Trondheim, Norway\\ E-mail: \email{lars.leganger@ntnu.no}}

\preprint{\arXivid{0903.4596}}

\date{\today}

\abstract{
We calculate the pressure of massless $\phi^4$-theory to order $g^8\log(g)$
at weak coupling.
The contributions to the pressure arise from the hard momentum scale of order
$T$ and the soft momentum scale of order $gT$.
Effective field theory methods and dimensional reduction are used to separate
the contributions from the two momentum scales: The hard contribution
can be calculated as a power series in $g^2$ using naive perturbation
theory with bare propagators. The soft contribution can be calculated
using an effective theory in three dimensions, whose coefficients
are power series in $g^2$. This contribution is a power series
in $g$ starting at order $g^3$. The calculation of the hard part
to order $g^6$ involves a complicated four-loop sum-integral that was recently
calculated by Gynther, Laine, Schr\"oder, Torrero, and Vuorinen.
The calculation of the
soft part requires calculating the mass parameter in the effective theory
to order $g^6$ and 
the evaluation of five-loop vacuum diagrams in three
dimensions. This gives the free energy correct up to order $g^7$.
The coefficients of the effective theory satisfy a set of renormalization group
equations that can be used to sum up leading and subleading logarithms
of $T/gT$. We use the solutions to these equations to obtain a result for
the free energy which is correct to order $g^8\log(g)$.
Finally, we investigate the convergence of the perturbative series.}

\begin{document}

\section{Introduction}
In recent years there has been significant progress in our understanding
of thermal field theories in equilibrium ~\cite{Blaizot:2003tw,Rischke:2003mt,Kraemmer:2003gd,Andersen:2004fp}. 
Part of the progress is based on the 
developement of the 
calculational technology necessary to perform loop calculations beyond 
the first correction.
The motivation to carry out such difficult higher-order calculations of e.g.
the pressure in thermal QCD is its relevance to heavy-ion collisions and
the early universe. The pressure in nonabelian gauge theories
has been calculated perturbatively through order $g^4$ in Ref.~\cite{Arnold:1994ps,Arnold:1994eb},
to order $g^5$ in Refs.~\cite{Zhai:1995ac,Braaten:1996rq}, and to order
$g^6\log(g)$ in Ref.~\cite{Kajantie:2003ax}.  There are three momentum scales that
contribute to the pressure in thermal QCD - hard momenta of order $T$,
soft momenta of order $gT$, and supersoft momenta of order $g^2T$.
The next order -- order $g^6$ -- is the first order at which all three
momentum
scales contribute to the pressure
and it is also the order at which perturbation theory breaks down due to
infrared divergences~\cite{Linde:1980ts,Gross:1980br}.
The pressure contains a nonperturbative contribution from the supersoft scale
that can be estimated numerically~\cite{Hietanen:2004ew,Hietanen:2006rc,DiRenzo:2006nh}.
It also contains a presently unknown contribution from the hard scale.
This contribution can be calculated by evaluating highly nontrivial 
four-loop vacuum diagrams with unresummed propagators. 
As a step in this direction, Gynther, Laine, Schr\"oder, Torrero, and Vuorinen
considered the simpler problem of $\phi^4$-theory at finite temperature
and calculated the free energy to order $g^6$~\cite{Gynther:2007bw}. 
A difficult part of
the calculation was to evaluate the four-loop triangle sum-integral, using
the techniques developed by Arnold and Zhai in Refs.~\cite{Arnold:1994ps,Arnold:1994eb}.

In hot field theories at weak coupling, the momentum scales in the plasma
are well separated and it is advantageous
to use effective field theory methods to organize the calculations of the
pressure into separate contributions from the hard, soft and supersoft scales.
The basic idea is that the mass of the nonzero Matsubara modes are of order $T$
and heavy. Since these modes are heavy, they decouple from the light modes, 
i.e. the static Matsubara modes.
In particular, all fermionic modes decouple since their masses are always of order
$T$.
The contributions from the nonzero Matsubara modes
to thermodynamic quantities can be
calculated using bare propagators and are encoded in the parameters of the
effective theory. 
Integrating out the hard scale $T$, i.e. integrating out the nonzero
Matsubara frequencies, 
leaves us with an effective dimensionally
reduced theory for the scales $gT$ and $g^2T$~\cite{Braaten:1996rq}.
In the case of QCD, the effective theory 
is an $SU(N)$ gauge theory coupled to an adjoint Higgs.
The process is known as dimensional 
reduction~\cite{Ginsparg:1980ef,Appelquist:1981vg,Landsman:1989be,Braaten:1995cm,Kajantie:1995dw}.
The next step is to construct a second effective theory 
for the scale $g^2T$ by integrating out
the scale $gT$ from the problem~\cite{Braaten:1996rq}. 
It amounts to integrating out the
adjoint Higgs and this step can also be made in perturbation
theory. This effective theory is a nonabelian gauge theory in three dimensions,
which is confining with a nonperturbative mass gap of order 
$g^2T$~\cite{Gross:1980br}.
This theory must be treated nonperturbatively and gives the nonperturbative
contribution to the pressure mentioned above.

In the present paper we consider the thermodynamics of massless $\phi^4$-theory
and calculate the pressure through order $g^8\log(g)$ 
in a weak-coupling expansion
using effective field theory. Calculations in scalar field theory are
simplified by the fact that the supersoft scale $g^2T$ does not appear
and so we  only need to construct a single effective theory for the 
soft scale $gT$.
This theory is infrared safe to all orders in perturbation
theory due to the generation of a thermal mass of order $gT$.
Compared to the $g^6$-calculations of Ref.~\cite{Gynther:2007bw}, 
the next order requires
the matching of the mass parameter to three loops and the evaluation of
some five-loop vacuum diagrams in the effective theory.
The matching involves a nontrivial three-loop sum-integral that was
calculated recently in Ref.~\cite{Andersen:2008bz}.

The paper is organized as follows. In Sec.~II, we briefly discuss effective
field theory 
and determine the coefficients of the dimensionally reduced theory.
In Sec.~III, we use the effective theory and calculate the soft contributions
to the pressure. In Sec.~IV, we present and discuss
our final results for the pressure.  
In Sec.~V, we 
summarize. In Appendix A and B, we list the necessary sum-integrals and 
integrals. In Appendix C, we calculate explicitly some of the new
three-dimensional integrals that we need.

\section{Effective field theory}
In this section, we briefly discuss the three-dimensional
effective field theory and the
matching procedure used to determine its coefficients. For a detailed
discussion, see e.g. Refs.~\cite{Braaten:1995cm,Kajantie:1995dw}.

The Euclidean
Lagrangian density for a massless scalar field with a $\Phi^4$-interaction 
is
\bqa
\label{ori}
{\cal L}={1\over2}(\partial_{\mu}\Phi)^2
+{g^2\over24}\Phi^4+\Delta{\cal L}\;,
\eqa
where $g$ is the coupling constant and $\Delta {\cal L}$ 
includes counterterms. This term reads
\bqa
\Delta {\cal L}&=&{1\over2}\Delta Z_{\Phi}(\partial_{\mu}\Phi)^2
+{1\over24}\Delta g^2\Phi^4\;.
\eqa
In the present case we need the counterterm $\Delta g^2$
to 
next-to-leading
order i $g^2$. It is given by
\bqa
\Delta g^2&=&\left[{3\over2\epsilon}\alpha
+\left({9\over4\epsilon^2}-{17\over12\epsilon}\right)\alpha^2\right]g^2\;,
\label{gcount}
\eqa
where $\alpha=g^2/(4\pi)^2$.
We denote by $\phi(x)$ the field in the effective theory. It can be
approximately, 
i.e. up to field redefinitions, be identified with zero-frequency
mode of the field $\Phi$ in the original theory.
The Lagrangian of the effective theory can be then be written as
\bqa
{\cal L}_{\rm eff}&=&{1\over2}(\nabla\phi)^2
+{1\over2}m^2\phi^2+{g^2_3\over24}\phi^4+...\;,
\label{lageff}
\eqa
where $m$ is the mass of the theory and $g_3^2$ is the quartic coupling.
The dots indicate an infinite series of higher-order operators
consistent with the symmetries, such as rotational invariance and
the discrete symmetry $\phi\rightarrow-\phi$.
In Eq.~(\ref{lageff}), we have omitted a coefficient $f$ of the unit operator.
Its interpretation
is that it gives the contribution to the free energy from the hard scale $T$.

For the calculation of the pressure to order $g^8\log(g)$, we need to know
$f$ and 
the mass parameter $m^2$ to order $g^6$ and the coupling constant $g_3^2$
to order $g^4$, i.e. we consider $\phi^4$-theory in three spatial 
dimensions~\footnote{Power counting tells one that the operator
$(\phi\nabla\phi)^2$ contributes to the free energy first at order $g^8$.}. 
This theory is superrenormalizable and only the mass
needs renormalization~\cite{Farakos:1994kx}.
The parameters in the effective Lagrangian~(\ref{lageff})
are determined by calculating static correlation functions in the two
theories at long distances $R$, i.e. $R\gg1/T$, and demanding that they be the 
same~\cite{Braaten:1995cm}. 
In the matching calculations, we are employing 
\emph{strict perturbation theory}~\cite{Braaten:1995cm}.
This amounts to doing 
perturbative calculations in power series in $g^2$ in which we
treat the mass parameter as a perturbation in the effective theory.
The Lagrangian is therefore split into a free and an interacting part
according to
\bqa
{\cal L}_{\rm eff}^{\rm free}&=&
{1\over2}(\nabla\phi)^2\;, \\
{\cal L}_{\rm eff}^{\rm int}&=&
{1\over2}m^2\phi^2+{g^2_3\over24}\phi^4+...\;.
\eqa
Strict perturbation theory 
gives rise to infrared divergences in the calculation
that physically are cut off by the generation of a thermal mass $m$. 
The same infrared divergences appear in the loops in the full theory and
so they cancel in the matching calculations.
The incorrect treatment of the infrared divergences and the physics
on the scale $gT$ is not problematic since this will be taken care of
by calculations in the effective theory. The matching calculations 
treat the physics on the hard scale correctly and the physics
on that scale is encoded in the parameters of the three-dimensional
effective Lagrangian.

However, the matching calculations of the parameters in 
${\cal L}_{\rm eff}$ are complicated by ultraviolet divergences.
Those divergences that are associated with the full four-dimensional
theory are removed by renormalization of the coupling constant $g$.
The remaining divergences are cancelled by the extra counterterms that are
determined by the ultraviolet divergences in the effective theory.
These divergences are regulated by introducing a cutoff $\Lambda$.
The cutoff $\Lambda$ can be thought of as an arbitrary factorization scale 
that separates the scale $T$ from the scale $gT$ (or smaller) 
which can be treated in the
effective theory~\cite{Braaten:1995cm}. 
The parameters in the effective theory therefore
depend on the cutoff $\Lambda$ in order to cancel the 
$\Lambda$-dependence of the loop integrals in the effective theory.

\subsection{Coupling constant}
\FIGURE{\includegraphics[width=3.6cm]{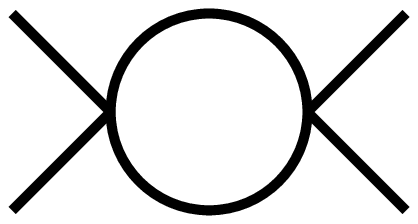}\caption{One-loop Feynman graph that contributes to the coupling $g_3^2$ in the effective theory.}\label{g3}} 
To leading order in the coupling $g^2$, we can simply read off
the coupling $g_3^2$ from the Lagrangian of the full theory.
Making the replacement $\Phi\rightarrow \sqrt{T}\phi$
in the Lagrangian~(\ref{ori}) and comparing 
$\int_0^{\beta}d\tau\,{\cal L}$ with ${\cal L}_{\rm eff}$, 
we conclude that $g_3^2=g^2T$.
The one-loop graph needed for the matching of the coupling $g_3^2$
to next-to-leading order in $g^2$
is shown in Fig.~\ref{g3}. Since the 
loop correction vanishes in the effective theory due to the fact that we are
using massless propagators,
the matching equation reduces to
\bqa
g_3^2&=&g^2T-{3\over2}g^4T\sumint_P{1\over P^4}+\Delta_1g^2T\;,
\eqa
where $\Delta_1g^2$ is the order-$g^4$ coupling constant counterterm in
Eq.~(\ref{gcount}).
After renormalization, we find
\bqa
g_3^2(\Lambda)&=&g^2(\mu)T\left[1-{3g^2\over{(4\pi)^2}}
\left(\log{\mu\over4\pi T}+\gamma_E\right)-\right.\nonumber\\
&&\left.-{3g^2\over{(4\pi)^2}}
\left(\log^2{\mu\over4\pi T}+2\gamma_E\log{\mu\over4\pi T}
+{\pi^2\over8}-2\gamma_1\right)\epsilon
\right]\;,
\label{g3final}
\eqa
where $g^2=g^2(\mu)$ is the coupling constant at the scale $\mu$ in the 
$\rm\overline{MS}$ scheme and we have
kept the order-$\epsilon$ terms in $g_3^2$ for later use.
We have used the renormalization group equation 
for the running coupling constant $g^2$,
\bqa
\mu{\partial\over\partial\mu}\alpha&=&
3\alpha^2-{17\over3}\alpha^3\;,
\label{arun}
\eqa
to change the scale from $\Lambda$ to $\mu$. The right-hand side 
of Eq.~(\ref{g3final}) is
independent of $\Lambda$. In fact, since 
the coupling $g_3^2$ does not require
renormalization in three dimensions, it
satisfies the renormalization group equation
\bqa
\Lambda{\partial\over\partial\Lambda}g_3&=&
0
\;.
\eqa

\subsection{Coefficient of unit operator}
The partition function in the full theory is given by the path integral
\bqa
Z&=&\int{\cal D}\Phi\,e^{-\int_0^{\beta}d\tau\int d^3x {\cal L}}\;,
\eqa
and the pressure is then given by ${\cal P}=T\log {\cal Z}/V$, 
where $V$ is the volume of the 
system.
In terms of the effective theory, the partition function can be written as
\bqa
{\cal Z}&=&e^{-fV}\int{\cal D}\phi\,e^{-\int d^3x {\cal L}_{\rm eff}}\;.
\eqa
The matching then yields
\bqa
\log{\cal Z}&=&-fV+\log{\cal Z}_{\rm eff}\;,
\label{matchf}
\eqa
where ${\cal Z}_{\rm eff}$ is the partitition function of the three-dimensional
theory. Equivalently, we can  write 
${\cal F}={\cal F}_{\rm hard}+{\cal F}_{\rm soft}$, where ${\cal F}_{\rm hard}=fT$
and ${\cal F}_{\rm soft}=-T\log{\cal Z}_{\rm eff}/V$.
Now since calculations in strict perturbation theory in the 
effective theory is carried out using bare propagators, there is no
scale in the vacuum graphs. This implies that they vanish in dimensional
regularization and that $\log{\cal Z}_{\rm eff}=0$. 
Eq.~(\ref{matchf}) then tells us that $f$ is given by a strict loop expansion
in four dimensions.

The vacuum diagrams through four loops
are shown in 
Figs.~\ref{vacuum1}--\ref{vacuum4}.





\DOUBLEFIGURE{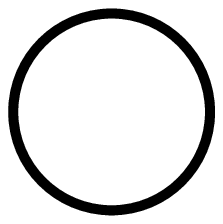,width=1.4cm}{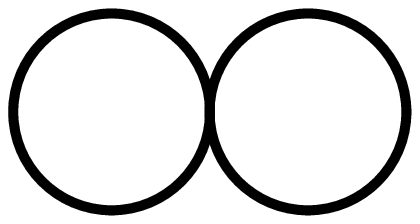,width=3.0cm}{One-loop vacuum diagram.\label{vacuum1}}{Two-loop vacuum diagram.\label{vacuum2}}
\DOUBLEFIGURE{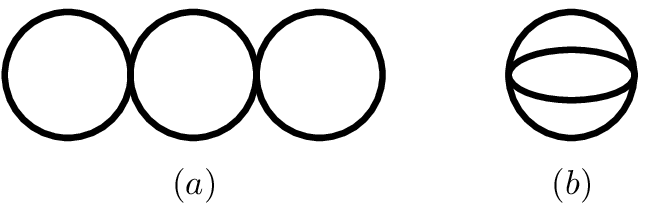,width=6cm}{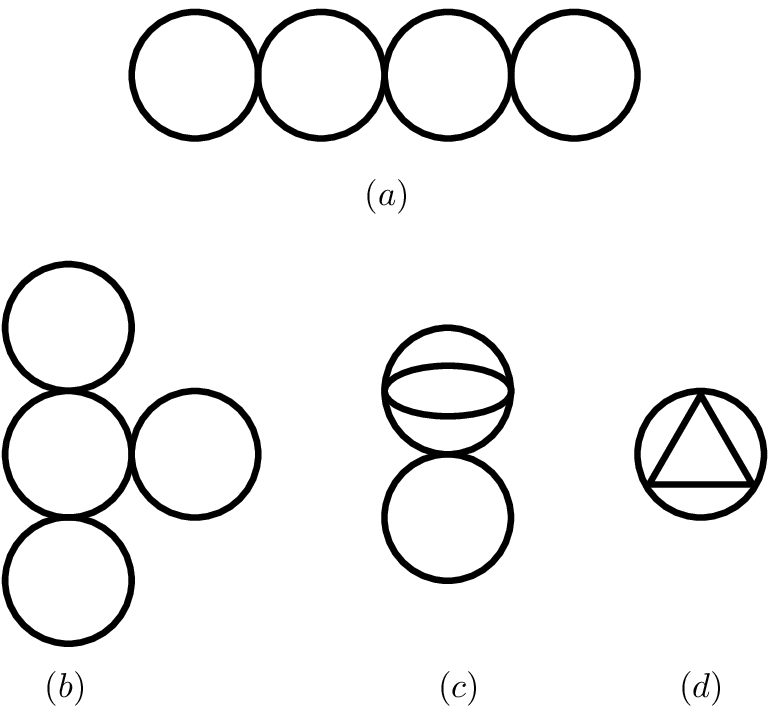,width=6cm}{Three-loop vacuum diagrams.\label{vacuum3}}{Four-loop vacuum diagrams.\label{vacuum4}}

We can then write
\bqa\nonumber
{\cal F}_{\rm hard}&=&{\cal F}_{\rm 0}^{(h)}+
{\cal F}_{\rm 1}^{(h)}+{\cal F}_{\rm 2a}^{(h)}+{\cal F}_{\rm 2b}^{(h)}
+{\cal F}_{\rm 3a}^{(h)}+{\cal F}_{\rm 3b}^{(h)}
+{\cal F}_{\rm 3c}^{(h)}+{\cal F}_{\rm 3d}^{(h)}+\\
&&+{{\cal F}_{\rm 1}^{(h)}\over g^2}(\Delta_1g^2+\Delta_2g^2)
+2\left({{\cal F}_{\rm 2a}^{(h)}\over g^2}
+{{\cal F}_{\rm 2b}^{(h)}\over g^2}
\right)\Delta_1g^2\;,
\eqa
where $\Delta_1g^2$ and $\Delta_2g^2$
are the order-$g^4$ and order-$g^6$ coupling constant counterterms,
respectively, given in Eq.~(\ref{gcount}).
The superscript $h$ indicates that the expression gives the hard contribution
to the free energy.
The expressions for the diagrams are
\bqa
{\cal F}_{0}^{(h)}&=&{1\over2}\sumint_P\log P^2\;,\\
{\cal F}_{\rm 1}^{(h)}&=&
{1\over8}g^2\left(\sumint_P{1\over P^2}\right)^2\;,\\
{\cal F}_{\rm 2a}^{(h)}&=&
-{1\over16}g^4\left(\sumint_P{1\over P^2}\right)^2\sumint_Q{1\over Q^4}\;,\\
{\cal F}_{\rm 2b}^{(h)}&=&
-{1\over48}g^4\sumint_{PQR}{1\over P^2Q^2R^2(P+Q+R)^2}\;, \\
{\cal F}_{\rm 3a}^{(h)}&=&
{1\over32}g^6\left(\sumint_P{1\over P^2}\right)^2
\left(\sumint_Q{1\over Q^4}\right)^2\;, \\
{\cal F}_{\rm 3b}^{(h)}&=&
{1\over48}g^6\sumint_P{1\over Q^6}\left(\sumint_P{1\over P^2}\right)^3\;, \\
{\cal F}_{\rm 3c}^{(c)}&=&
{1\over24}g^6\sumint_P{1\over P^2}\sumint_{KQR}{1\over K^4Q^2R^2(K+Q+R)^2}\;,
\\ 
{\cal F}_{\rm 3d}^{(h)}&=&
{1\over48}g^6\sumint_P\left[\Pi(P)\right]^3\;,
\eqa
where the symbol $\sumint$ is defined in Eq.~(\ref{sumint-def})
and the self-energy $\Pi(P)$ is defined in Eq.~(\ref{pidef1}).
The expressions for the sum-integrals are listed in Appendix A.
After renormalization, the final expression is~\cite{Gynther:2007bw}
\bqa\nonumber
{\cal F}_{\rm hard}(\Lambda)&=&
-{\pi^2T^4\over90}\times \nonumber \\ &&\nonumber \times \bigg\{
1-{5\over4}\alpha
+{15\over4}\alpha^2
\left[
\log{\mu\over4\pi T}
+{1\over3}\gamma_E+{31\over45}+{4\over3}{\zeta^{\prime}(-1)\over\zeta(-1)}
-{2\over3}{\zeta^{\prime}(-3)\over\zeta(-3)}
\right]+
\\&& \nonumber
\hphantom{\times\bigg\{}
+{15\over16}\alpha^3\times\nonumber\\ && \hphantom{\times\bigg\{} \times
\left[
{\pi^2\over\epsilon}
-12\log^2{\mu\over4\pi T}
-\left({1084\over45}+8\gamma_E+
32{\zeta^{\prime}(-1)\over\zeta(-1)}
-16{\zeta^{\prime}(-3)\over\zeta(-3)}
\right)
\times \nonumber \right. \\ && \left. \hphantom{\times\bigg\{\times\bigg[}\times
\log{\mu\over4\pi T}
+8{\pi^2}\log{\Lambda\over4\pi T}
-{134\over9}-
{25\over3}\gamma_E^2-{1\over27}\zeta(3)
+{31\over15}\gamma_E
-\right. \nonumber\\ \nonumber&&\left.\hphantom{\times\bigg\{\times\bigg[}
-{\pi^2\over2}+4\gamma_E\pi^2
-{206\over9}{\zeta^{\prime}(-1)\over\zeta(-1)}
-{16\over3}\gamma_1+8\gamma_E{\zeta^{\prime}(-3)\over\zeta(-3)}
+\right. \\ \nonumber&&\hphantom{\times\bigg\{\times\bigg[}\left.
+{4\over3}\gamma_E{\zeta^{\prime}(-1)\over\zeta(-1)}
-8\left({\zeta^{\prime}(-1)\over\zeta(-1)}\right)^2
-{20\over3}{\zeta^{\prime\prime}(-1)\over\zeta(-1)}-\right. \\ &&\hphantom{\times\bigg\{\times\bigg[}\left.
-{2\over3}C_{\rm ball}^{\prime}+2C^a_{\rm triangle}+\pi^2C^b_{\rm triangle}
\right]
+{\cal O}(\epsilon)\bigg\}\;,
\label{fhard}
\eqa
where $\alpha=\alpha(\mu)$,
$C_{\rm ball}^{\prime}=48.7976$, $C^a_{\rm triangle}=-25.7055$, and
$C^b_{\rm triangle}=28.9250$. 
We have used the renormalization group equation for $g^2$ to change the
renormalization scale from $\Lambda$ to $\mu$.
Note that the final results contains a pole in $\epsilon$.
We cancel it by adding a counterterm $T\delta f_{\rm }$~\cite{Braaten:1996rq}.
The term $\delta f_{\rm}$ can be determined by calulating the 
ultraviolet divergences in the effective theory.
The triangle diagram in three dimensions has a logarithmic ultraviolet
divergence and the counterterm needed to cancel this divergence is given by
\bqa
\label{delta}
\delta f_{\rm }&=&{g_3^6\pi^2\over1536(4\pi)^4\epsilon}\;.
\eqa
If we express the counterterm in terms of the coupling $g$ of the full
theory, we must take into account that $g_3^6$ multiplies a pole in 
$\epsilon$ and it therefore picks up finite terms.
These terms will be of order $g^8$ and can be neglected in the present 
calculation~\footnote{Note that minimal subtraction in the full theory and
in the effective theory are not equivalent. The difference is the finite
terms mentioned above~\cite{Braaten:1996rq}.}. 
The coefficient $f$ satisfies
the evolution equation
\bqa
\Lambda{\partial\over\partial\Lambda}f&=&
-{\pi^2\over192(4\pi)^4}g_3^6\;.
\eqa
This follows from the scale dependence of the triangle diagram in three 
dimensions and the fact that the $\Lambda$-dependence of $f$
must cancel the scale dependence in the effective theory.

\subsection{Mass parameter}
The simplest way of determining the mass parameter $m^2$ 
is by matching the Debye or screening mass $m_D$ in the full theory and in the
effective theory~\cite{Braaten:1995cm}.
The Debye mass $m_D$ is given by the pole of static propagator, i.e. by 
\bqa
p^2+\tilde{\Pi}(p_0=0,p)&=&0\;,\qquad p^2=-m_D^2\;,
\label{pifull}
\eqa
where $\tilde{\Pi}(p_0,p)$ denotes the self-energy 
function.
In the effective
theory, the equation is 
\bqa
p^2+m^2+\Pi_{\rm eff}(p)&=&0\;,\qquad p^2=-m_D^2\;,
\label{pieff}
\eqa
where $\Pi_{\rm eff}(p)$ is the self-energy in the effective
theory. 
Since the self-energy in the full theory is expanded around $p=0$, we should
do to the same in the effective theory (see discussion below).
The loop integrals are therefore evaluated at zero external momentum and since
the matching is carried
out using massless propagators there is no scale in the loop integrals.
They therefore vanish in 
in dimensional regularization, i.e. 
$\Pi_{\rm eff}(0)=\Pi^{\prime}_{\rm eff}(0)=...=0$.
Using this fact and equating~(\ref{pifull}) and~(\ref{pieff}), we obtain
$m^2\approx m_D^2$~\footnote{Note that we use the symbol ``$\approx$''
to emphasize that the the mass parameter $m^2$ is equal to the 
Debye mass $m_D^2$ only in strict perturbation theory. The interpretation is
that $m$ gives the
contribution to the Debye mass from the hard scale $T$.}
\bqa
m^2_D&=&\tilde{\Pi}(p_0=0,p=im_D)\;.
\label{matchmaker}
\eqa
The diagrams that contribute to the self-energy $\tilde{\Pi}(P)$ 
through three loops are shown in Fig~\ref{self}.

\FIGURE{\includegraphics[width=10cm]{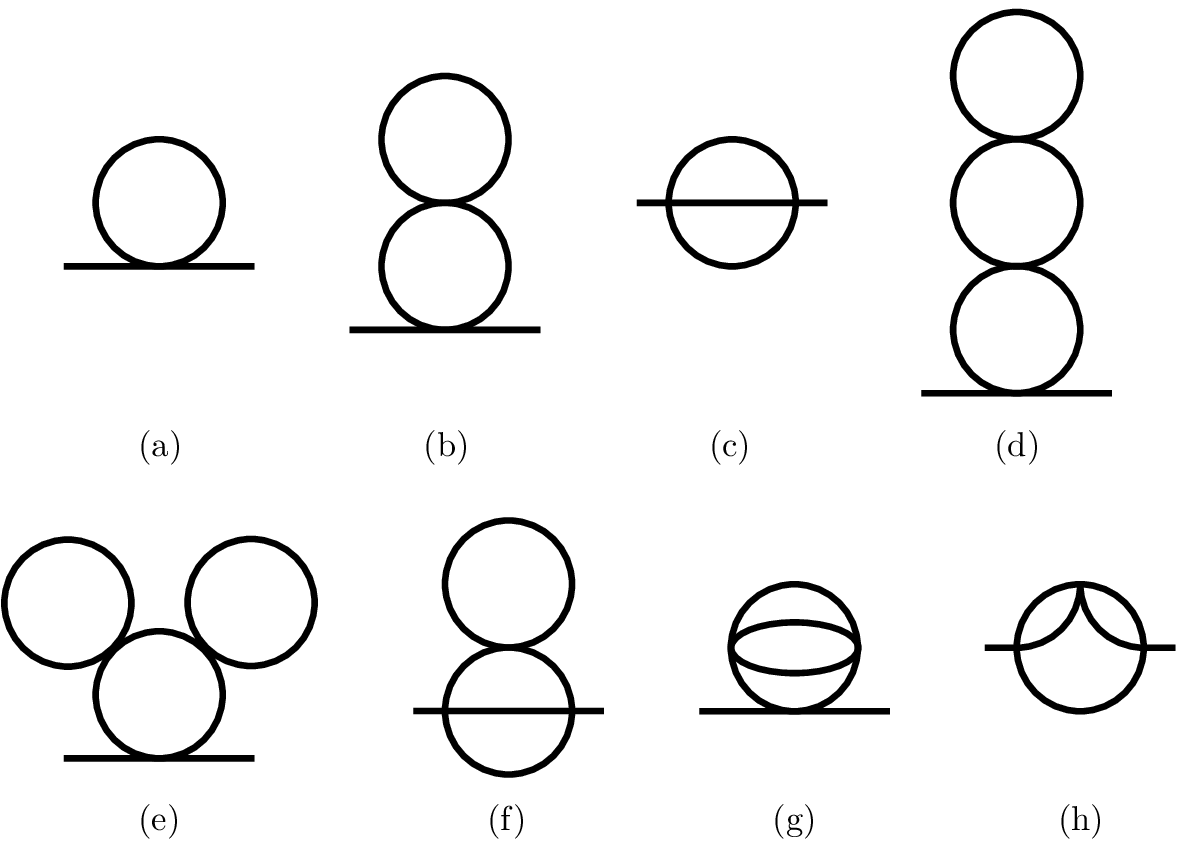}\caption{Feynman graphs that contribute to the self-energy through three loops.}\label{self}}

The self-energy $\tilde{\Pi}(P)$ is given 
by
\bqa
\tilde{\Pi}(P)&=&\tilde{\Pi}_{1}^{(h)}(P)+
\tilde{\Pi}_{2}^{(h)}(P)+\tilde{\Pi}_{3}^{(h)}(P)+\nonumber\\&&
+{\tilde{\Pi}_{1}^{(h)}(P)\over g^2}\left(\Delta_1g^2+\Delta_2g^2\right)
+2{\tilde{\Pi}_{2}^{(h)}(P)\over g^2}\Delta_1g^2
\;.
\eqa
The expression for the various terms in the self-energy are
given by 
\bqa
\tilde{\Pi}_{1}^{(h)}(P)&=& {1\over2}g^2\sumint_Q{1\over Q^2}\;,
\\
\tilde{\Pi}_{2a}^{(h)}(P)&=&
-{1\over4}g^4\sumint_{QR}{1\over Q^4R^2}\;,\\
\tilde{\Pi}_{2b}^{(h)}(P)&=&
-{1\over6}g^4\sumint_{QR}{1\over Q^2R^2(P+Q+R)^2}\;, \\
\tilde{\Pi}_{3a}^{(h)}(P)&=&
{1\over8}g^6\sumint_Q{1\over Q^2}\left(\sumint_R{1\over R^4}\right)^2
\;,\\
\tilde{\Pi}_{3b}^{(h)}(P)&=&
{1\over8}g^6\sumint_Q{1\over Q^6}\left(\sumint_R{1\over R^2}\right)^2
\,\\
\tilde{\Pi}_{3c}^{(h)}(P)&=&
{1\over4}g^6\sumint_{K}{1\over K^2}\sumint_{QR}{1\over Q^4R^2(P+Q+R)^2}
\;, \\
\tilde{\Pi}_{3d}^{(h)}(P)&=&{1\over12}g^6
\sumint_{KQR}{1\over K^4Q^2R^2(K+Q+R)^2}
\;, \\
\tilde{\Pi}_{3e}^{(h)}(P)&=&{1\over4}g^6\sumint_{Q}{1\over(P+Q)^2}[\Pi(Q)]^2
\;.
\eqa
Since the leading-order solution to Eq.~(\ref{matchmaker}) 
gives a value of $p$ that is of the order
$gT$, it is justified to expand the loop diagrams in a Taylor series
around $p=0$. We can then write Eq.~(\ref{matchmaker}) as
\bqa
m_D^2&=&\tilde{\Pi}_1^{(h)}(0)+\tilde{\Pi}_2^{(h)}(0)
+\tilde{\Pi}_2^{(h)\prime}(0)p^2
+\tilde{\Pi}_3(0)+...\;,
\hspace{0.2cm}p^2=-m_D^2\;,
\eqa
or $m_D^2=\tilde{\Pi}^{(h)}_1(0)+\tilde{\Pi}^{(h)}_2(0)+\tilde{\Pi}^{(h)}_3(0)
-\tilde{\Pi}_1(0)\tilde{\Pi}_2^{\prime}(0)$\;.
We then need 
the two-loop self-energy diagram 
$\tilde{\Pi}_{\rm 2b}(P)$ to order $p^2$, while 
the three-loop self-energy diagrams $\tilde{\Pi}_{\rm 3c}^{(h)}(P)$
and $\tilde{\Pi}_{\rm 3e}^{(h)}(P)$
can be evaluated at $p=0$.
This yields
\bqa
\label{alsowave}
\tilde{\Pi}_{\rm 2b}^{(h)}(P)&=&-{1\over6}g^4
\sumint_{QR}{1\over Q^2R^2(Q+R)^2}
-{1\over6}g^4p^2\sumint_{QR}{(4/d)q^2-Q^2\over Q^6R^2(Q+R)^2}
+{\cal O}(p^4)\;,\\
\tilde{\Pi}^{(h)}_{3c}(0)
&=&{1\over4}g^6\sumint_{K}{1\over K^2}\sumint_{QR}{1\over Q^4R^2(Q+R)^2}\;,
\\\tilde{\Pi}^{(h)}_{3e}(0)
&=&{1\over4}g^6\sumint_{Q}{1\over Q^2}[\Pi(Q)]^2\;.
\eqa
The sum-integrals needed are listed in Appendix A. After renormalization,
we obtain
\bqa
m^2(\Lambda)&=&{1\over24}g^2(\Lambda)T^2\times \nonumber \\ && \times \bigg\{
1+{g^2\over(4\pi)^2}\left[
{1\over\epsilon}+\log{\Lambda\over4\pi T}
+2-\gamma_E+2{\zeta^{\prime}(-1)\over\zeta(-1)}
\right]-{6g^4\over(4\pi)^4}\times \nonumber\\ \nonumber
&&\hphantom{\times\bigg\{}\times \left[{1\over\epsilon}\left(\log{\Lambda\over4\pi T}
+\gamma_E\right)
+{7\over2}\log^2{\Lambda\over4\pi T}
+\left({19\over18}+5\gamma_E+2{\zeta^{\prime}(-1)\over\zeta(-1)}
\right)\log{\Lambda\over4\pi T}+\right.\\ \nonumber&&\hphantom{\times\bigg\{\times\bigg[}\left.
+{2851\over864}
-{95\over48}\gamma_E^2
-{119\over144}\gamma_E
-{1\over144}\zeta(3)
-9\gamma_1
+{\zeta^{\prime}(-1)\over\zeta(-1)}
\left(
{113\over72}+{17\over12}\gamma_E\right)-
\right.\\ &&\hphantom{\times\bigg\{\times\bigg[}\left.
-
{1\over4}{\zeta^{\prime\prime}(-1)\over\zeta(-1)}
+{29\over32}\pi^2-2\gamma_E\log(2\pi)
+2\log^2(2\pi)
-{1\over24}C_{\rm ball}^{\prime}+{1\over4}C_{I}
\right] + \nonumber\\&&
\hphantom{\times\bigg\{}+{\cal O}(\epsilon)\bigg\}\;,
\label{m3prefinal}
\eqa
where 
$g=g(\Lambda)$
and 
$C_I=-38.4672$. The mass parameter through order $g^4$ 
is known to order $\epsilon$~\cite{Kajantie:2003ax}, but we only need it to order
$\epsilon^0$.
We notice that the mass parameter contains uncancelled poles in 
$\epsilon$.
It is advantageous to write the mass term as a sum of a finite piece
$\tilde{m}^2$
and a counterterm $\Delta m^2$, where
\bqa\nonumber
\tilde{m}^2(\Lambda)&=&{1\over24}g^2(\mu)T^2\times \nonumber\\&& \nonumber\times \bigg\{
1+{g^2\over(4\pi)^2}\left[
4\log{\Lambda\over4\pi T}-3\log{\mu\over4\pi T}
+2-\gamma_E+2{\zeta^{\prime}(-1)\over\zeta(-1)}
\right]
-\\ \nonumber&&\nonumber
\hphantom{\times\bigg\{}- {6g^4\over(4\pi)^4}\left[
4\log^2{\Lambda\over4\pi T}
-{3\over2}\log^2{\mu\over4\pi T}
+\left({19\over18}
-\gamma_E
+2{\zeta^{\prime}(-1)\over\zeta(-1)}
\right)\log{\mu\over4\pi T}
+\right. \\ \nonumber&&
\hphantom{\times\bigg\{- {6g^4\over(4\pi)^4}\bigg[}\left.
+4\gamma_E\log{\Lambda\over4\pi T}
+{2851\over864}-{95\over48}\gamma_E^2
-{119\over144}\gamma_E
-{1\over144}\zeta(3)
-7\gamma_1
+\right. \\ \nonumber&&
\hphantom{\times\bigg\{- {6g^4\over(4\pi)^4}\bigg[}\left.
+{\zeta^{\prime}(-1)\over\zeta(-1)}
\left(
{113\over72}+{17\over12}\gamma_E\right)
-
{1\over4}{\zeta^{\prime\prime}(-1)\over\zeta(-1)}
+{25\over32}\pi^2-2\gamma_E\log(2\pi)
+\right. \\&&
\hphantom{\times\bigg\{- {6g^4\over(4\pi)^4}\bigg[}\left.
+2\log^2(2\pi)
-{1\over24}C_{\rm ball}^{\prime}+{1\over4}C_{I}
\right]
+{\cal O}(\epsilon)\bigg\}\;,
\label{m3final}\\
\nonumber
\Delta m^2(\Lambda)&=&{g^4T^2\over24(4\pi)^2\epsilon}\left[
1-{6g^2\over{(4\pi)^2}}
\left(\log{\mu\over4\pi T}+\gamma_E\right) -\right.\\ \nonumber &&\hphantom{{g^4T^2\over24(4\pi)^2\epsilon}\bigg[}\left.
-{6g^2\over{(4\pi)^2}}
\left(\log^2{\mu\over4\pi T}+2\gamma_E\log{\mu\over4\pi T}
+{\pi^2\over8}-2\gamma_1\right)\epsilon
\right]\;,\\ 
&=&
{g^4_3(\Lambda)\over24(4\pi)^2\epsilon}\;,
\label{dm1}
\eqa
where $g=g(\mu)$
and we have used Eq.~(\ref{arun}) to change the renormalization scale
from $\Lambda$ to $\mu$.
The term $\Delta m^2$ acts as a counterterm in the effective theory. In fact,
the sunset diagram in three dimensions that contribute to the
self-energy is logarithmically divergent, whose divergence exactly is 
given by the right-hand side of Eq.~(\ref{dm1})~\cite{Farakos:1994kx}.
The mass parameter $\tilde{m}^²$ in three dimensions
therefore satisfies the evolution 
equation
\bqa
\Lambda{\partial\over\partial\Lambda}
\tilde{m}^2&=&{1\over6}{g_3^4\over(4\pi)^2}\;.
\eqa
In the remainder of the paper, we will use $m$ instead of
$\tilde{m}$ for covenience.

\section{Soft contributions}
In this section, we calculate the soft contributions
${\cal P}_{\rm soft}$ to the pressure.
This requires the calculations of vacuum diagrams in the 
effective theory~(\ref{lageff})
through five loops. In order to take into account the soft scale $gT$, 
we now include the mass term $m^2$ in the free part of the Lagrangian
and only the
quartic term in Eq.~(\ref{lageff}) is treated as an interaction.
The inclusion of the mass term in the propagators cuts off the infrared 
divergences that plagues naive perturbation theory in the full theory.

The one-loop vacuum diagram is shown in Fig~\ref{vacuum1}.
Its contribution to the free energy is given by 
\bqa
{\cal F}_0^{(s)}&=&{1\over2}T\int_p\log\left(p^2+m^2\right)\;,
\eqa
where the superscript $(s)$ indicates that the expression gives the
soft contribution to the free energy.
Using the expression in the Appendix B, we obtain
\bqa
{\cal F}_0^{(s)}&=&-{m^3T\over12\pi}\;.
\label{f0s}
\eqa
The two-loop vacuum diagram is shown in Fig~\ref{vacuum2}.
Its contribution to the free energy is given by
\bqa
{\cal F}_1^{(s)}&=&{1\over8}g_3^2T\left(\int_p{1\over p^2+m^2}\right)^2\;.
\label{f1s}
\eqa
Using the expression in the Appendix B, we obtain
\bqa
{\cal F}_1^{(s)}&=&{g_3^2m^2T\over8(4\pi)^2}\;.
\eqa
The three-loop vacuum diagrams are shown in Fig~\ref{vacuum3}.
The contribution to the free energy is given by
\bqa
{\cal F}_2^{(s)}
&=&{\cal F}_{\rm 2a}^{(s)}+{\cal F}_{\rm 2b}^{(s)}
+{\partial{\cal F}_{\rm 0}^{(s)}\over\partial m^2}\Delta m^2\,,
\eqa
where $\Delta m^2$ is the mass counterterm~(\ref{dm1}) 
in the effective theory and
\bqa
{\cal F}_{\rm 2a}^{(s)}
&=&-{1\over16}g_3^4T\left(\int_p{1\over p^2+m^2}\right)^2
\int_q{1\over(q^2+m^2)^2}\,,\\
{\cal F}_{\rm 2b}^{(s)}&=&
-{1\over48}g_3^4T\int_{pqr}{1\over p^2+m^2}{1\over q^2+m^2}{1\over r^2+m^2}
{1\over({\bf p}+{\bf q}+{\bf r})^2+m^2}
\;.
\eqa
Using the expression in the Appendix B, we obtain
\bqa
{\cal F}_2^{(s)}&=&{g_3^4mT\over96(4\pi)^3}\left[8\log{\Lambda\over2m}
+9-8\log2\right]\;.
\label{f2s}
\eqa
We note that all poles in $\epsilon$ cancel as they must since there are
no divergences from the hard part proportional to $g^4_3m$.

The four-loop vacuum diagrams are shown in Fig~\ref{vacuum4}.
The contribution to the free energy is given by
\bqa
{\cal F}_3^{(s)}
&=&{\cal F}_{\rm 3a}^{(s)}+{\cal F}_{\rm 3b}^{(s)}
+{\cal F}_{\rm 3c}^{(s)}+{\cal F}_{\rm 3d}^{(s)}
+{\partial{\cal F}_{\rm 1}^{(s)}\over\partial m^2}\Delta m^2\;,
\eqa
where the expressions for the diagrams are
\bqa
{\cal F}_{\rm 3a}^{(s)}&=&{1\over32}
g_3^6T
\left(\int_p{1\over p^2+m^2}\right)^2
\left(\int_q{1\over\left(q^2+m^2\right)^2}\right)^2 \;,\\
{\cal F}_{\rm 3b}^{(s)}&=&{1\over48}g_3^6T
\left(\int_p{1\over p^2+m^2}\right)^3
\int_q{1\over\left(q^2+m^2\right)^3} \;,\\
{\cal F}_{\rm 3c}^{(s)}&=&{1\over24}g_3^6T
\int_{pqr}{1\over(p^2+m^2)^2}{1\over q^2+m^2}{1\over r^2+m^2}
{1\over({\bf p}+{\bf q}+{\bf r})^2+m^2}\times\nonumber\\ && \times
\int_s{1\over s^2+m^2} \;, \\
{\cal F}_{\rm 3d}^{(s)}&=&{1\over48}g_3^6T
\int_{pqrs}{1\over q^2+m^2}{1\over({\bf p}+{\bf q})^2+m^2}
{1\over r^2+m^2}{1\over({\bf p}+{\bf r})^2+m^2}
\times\nonumber\\ && \hphantom{{1\over48}g_3^6T
\int_{pqrs}}\times
{1\over s^2+m^2}{1\over({\bf p}+{\bf s})^2+m^2}
\label{f3d}
\;.
\eqa
Using the expressions in the Appendix B, we obtain
\bqa
{\cal F}_3^{(s)}&=&{g_3^6T\over768(4\pi)^4}
\left[-4\left(4-\pi^2\right)\log{\Lambda\over2m}
-4+16\log2-42\zeta(3)+\pi^2(1+2\log2)
\right]
+\nonumber\\ && +{g_3^6T\pi^2\over1536(4\pi)^4\epsilon}
\;.
\label{f3s}
\eqa
The pole in $\epsilon$ in Eq.~(\ref{f3s}) arises from the 
triangle diagram in Eq.~(\ref{f3d}). This pole is cancelled by the counterterm
in Eq.~(\ref{delta}).


\EPSFIGURE{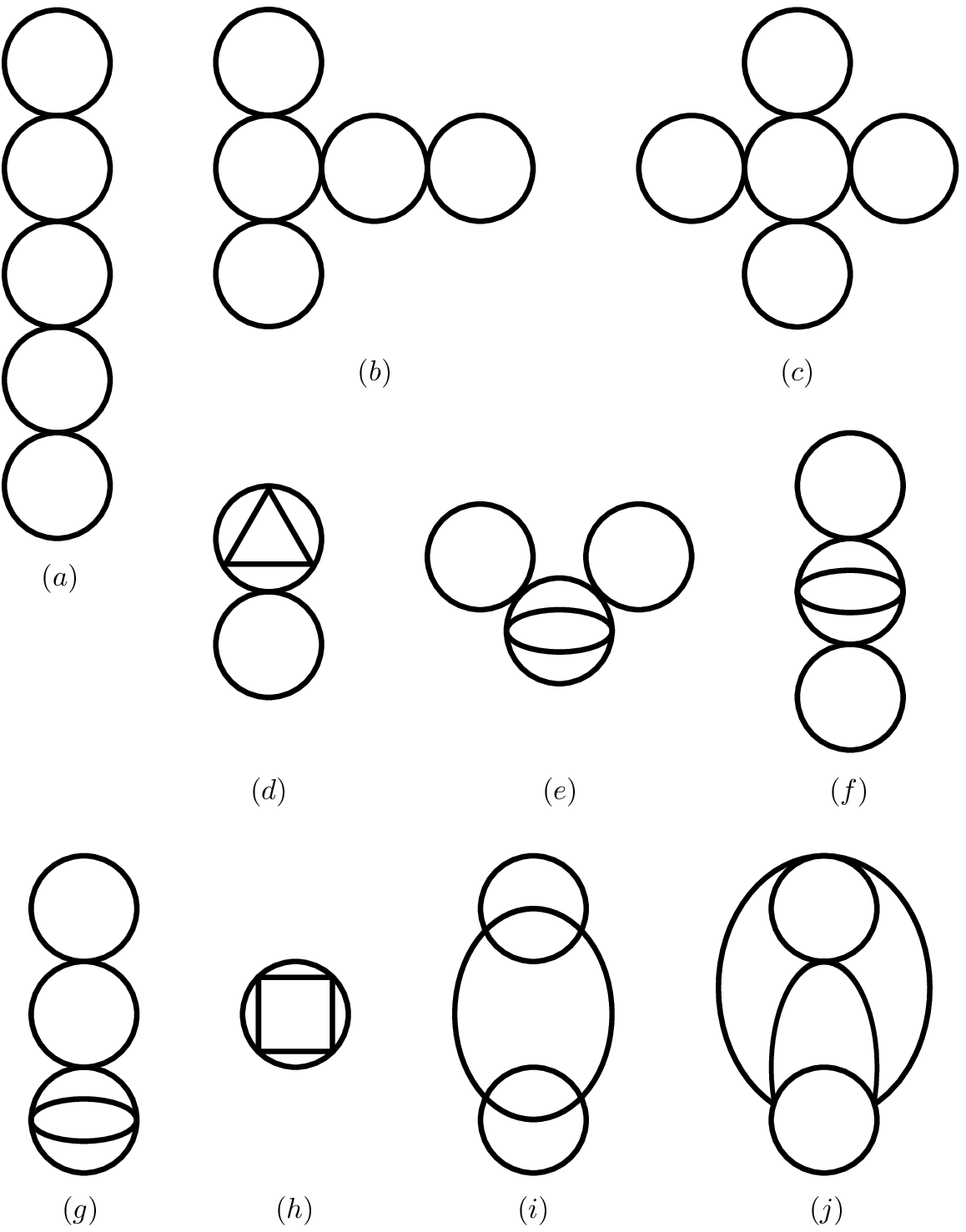,width=10cm}{Five-loop vacuum diagrams that contribute to the soft part of the free energy.\label{vacuum5}}

The five-loop vacuum diagrams are shown in Fig~\ref{vacuum5}.
The contributions to the free energy are given by
\bqa\nonumber
{\cal F}_4^{(s)}
&=&{\cal F}_{\rm 4a}^{(s)}+{\cal F}_{\rm 4b}^{(s)}
+{\cal F}_{\rm 4c}^{(s)}+{\cal F}_{\rm 4d}^{(s)}
+{\cal F}_{\rm 4e}^{(s)}+{\cal F}_{\rm 4f}^{(s)}
+{\cal F}_{\rm 4g}^{(s)}+{\cal F}_{\rm 4h}^{(s)}
+{\cal F}_{\rm 4i}^{(s)}+{\cal F}_{\rm 4j}^{(s)}
+\nonumber \\ &&
+{\partial{\cal F}_{\rm 2}^{(s)}\over\partial m^2}\Delta m^2
+{1\over2}{\partial^2{\cal F}_{\rm 0}^{(s)}\over(\partial m^2)^2}(\Delta m^2)^2
\;.
\eqa
where the expressions for the diagrams are
\bqa
{\cal F}_{\rm 4a}^{(s)}&=&-{1\over64}g_3^8T
\left(\int_p{1\over p^2+m^2}\right)^2
\left(\int_q{1\over\left(q^2+m^2\right)^2}\right)^3 \;,\\
{\cal F}_{\rm 4b}^{(s)}&=&-{1\over32}g_3^8T
\left(\int_p{1\over p^2+m^2}\right)^3
\int_q{1\over\left(q^2+m^2\right)^2} 
\int_r{1\over\left(r^2+m^2\right)^3} 
\;,\\
{\cal F}_{\rm 4c}^{(s)}&=&-{1\over128}g_3^8T
\left(\int_p{1\over p^2+m^2}\right)^4
\int_q{1\over\left(q^2+m^2\right)^4} \;,\\
{\cal F}_{\rm 4d}^{(s)}&=&-{1\over16}g_3^8T
\int_{pqrs}{1\over(q^2+m^2)^2}{1\over({\bf p}+{\bf q})^2+m^2}
{1\over r^2+m^2}{1\over({\bf p}+{\bf r})^2+m^2}
\times\nonumber\\&&\hphantom{-{1\over16}g_3^8T
\int_{pqrs}}\times
{1\over s^2+m^2}{1\over({\bf p}+{\bf s})^2+m^2}
\int_t{1\over t^2+m^2}
\;,\\
{\cal F}_{\rm 4e}^{(s)}&=&-{1\over48}g_3^8T
\int_{pqr}{1\over(p^2+m^2)^3}{1\over q^2+m^2}{1\over r^2+m^2}
{1\over({\bf p}+{\bf q}+{\bf r})^2+m^2}
\times\nonumber\\&&\times
\left(\int_s{1\over s^2+m^2}\right)^2
\;,
\eqa
\bqa
{\cal F}_{\rm 4f}^{(s)}&=&-{1\over32}g_3^8T
\int_{pqr}{1\over(p^2+m^2)^2}{1\over(q^2+m^2)^2}{1\over r^2+m^2}
{1\over({\bf p}+{\bf q}+{\bf r})^2+m^2}
\times\nonumber\\&&\times
\left(\int_s{1\over s^2+m^2}\right)^2
\;,\\
{\cal F}_{\rm 4g}^{(s)}&=&-{1\over48}g_3^8T
\int_{pqr}{1\over(p^2+m^2)^2}{1\over q^2+m^2}{1\over r^2+m^2}
{1\over({\bf p}+{\bf q}+{\bf r})^2+m^2}
\times\nonumber\\&&\times
\int_s{1\over s^2+m^2} \int_t{1\over(t^2+m^2)^2} 
\;,\\ \nonumber
{\cal F}_{\rm 4h}^{(s)}&=&-{1\over128}g_3^8T
\int_{pqrst}{1\over q^2+m^2}{1\over({\bf p}+{\bf q})^2+m^2}
{1\over r^2+m^2}{1\over({\bf p}+{\bf r})^2+m^2}
\times\nonumber\\&&\hphantom{-{1\over128}g_3^8T
\int_{pqrst}}\times
{1\over s^2+m^2}{1\over({\bf p}+{\bf s})^2+m^2}
{1\over t^2+m^2}{1\over({\bf p}+{\bf t})^2+m^2}
\;,\\
{\cal F}_{\rm 4i}^{(s)}&=&-{1\over144}g_3^8T
\int_p{1\over(p^2+m^2)^2}
\int_{qr}{1\over q^2+m^2}{1\over r^2+m^2}{1\over({\bf p}+{\bf q}+{\bf r})^2+m^2}
\times\nonumber\\&&\times
\int_{st}{1\over s^2+m^2}{1\over t^2+m^2}{1\over({\bf p}+{\bf s}+{\bf t})^2+m^2}
\;,\\
\nonumber
{\cal F}_{\rm 4j}^{(s)}&=&-{1\over32}g_3^8T
\int_{pqrst}{1\over q^2+m^2}{1\over({\bf p}+{\bf q})^2+m^2}
{1\over({\bf p}+{\bf r})^2+m^2}{1\over({\bf t}+{\bf r})^2+m^2}
{1\over r^2+m^2}
\times\nonumber\\&&\hphantom{-{1\over32}g_3^8T
\int_{pqrst}}\times
{1\over({\bf p}+{\bf s})^2+m^2}{1\over({\bf s}+{\bf t})^2+m^2}
{1\over s^2+m^2}
\;.
\eqa
Using the expressions in the Appendix B, we obtain
\bqa\nonumber
{\cal F}_{4}^{(s)}&=&-{g_3^8T\over288m(4\pi)^5}\times \nonumber \\ && \times
\left[\log^2{\Lambda\over2m}+{1\over4}\left(1-8\log2\right)\log{\Lambda\over2m}
-{15\over64}-{3\over8}\pi^2+{9\over8}\pi^2\log2
+\right. \nonumber\\ && \hphantom{\times\bigg[}\left.
+{23\over4}\log2+6\log^22
-6\log3-{81\over16}\zeta(3)
+5\rm{Li}_2(\mbox{$1\over4$})
+9C_{4j}
\right]\;,
\label{f4s}
\eqa
where $C_{4j}=0.443166$.
We note that all poles in $\epsilon$ cancel as they must since there are
no divergences from the hard part proportional to $g^8_3/m$.
Adding Eqs.~(\ref{f0s}), ~(\ref{f1s}), ~(\ref{f2s}), ~(\ref{f3s}), 
and~(\ref{f4s}) as well as the counterterm Eq.~(\ref{delta}), we 
obtain the soft contribution to the free energy
through five loops
\bqa\nonumber
{\cal F}_{0+1+2+3+4}^{(s)}&=&
-{m^3T\over12\pi}+{g_3^2m^2T\over8(4\pi)^2}
+{g_3^4mT\over96(4\pi)^3}\left[8\log{\Lambda\over2m}+9-8\log2\right]
+\\ &&\nonumber
+{g_3^6T\over768(4\pi)^4}\times \nonumber\\ &&\times\left[-4(4-\pi^2)\log{\Lambda\over2m}
-4+16\log2-42\zeta(3)+\pi^2(1+2\log2)
\right]
-\nonumber\\ && \nonumber
-{g_3^8T\over288m(4\pi)^5}\times \nonumber\\ &&\times
\left[\log^2{\Lambda\over2m}
+{1\over4}\left(1-8\log2\right)\log{\Lambda\over2m}
-{15\over64}-{3\over8}\pi^2+{9\over8}\pi^2\log2
+\right.\nonumber\\
&&\hphantom{\times\bigg[}\left.
+{23\over4}\log2+6\log^22
-6\log3-{81\over16}\zeta(3)
+5\rm{Li}_2(\mbox{$1\over4$})
+9C_{4j}
\right]\;.
\label{s15}
\eqa
Using the evolution equations for $g_3^2$ and $m^2$, it easy
to check that the free energy, Eq.~(\ref{fhard}) 
plus Eq.~(\ref{s15}) is independent of the factorization scale $\Lambda$.

By expanding the coupling $g_3^2$~(\ref{g3final}) 
and the mass parameter $m^2$~(\ref{m3final}) to the appropriate orders
in the various terms in~(\ref{s15}), we obtain the soft contribution 
through order $g^7$. This yields
\bqa\nonumber
{\cal F}_{\rm soft}^{}&=&
-{\pi^2T^4\over90}\times \nonumber\\ && \times \bigg\{{5\sqrt{6}\over3}{\alpha^{3/2}}
-{15\over2}{\alpha^2}-{15\sqrt{6}\over2}{\alpha^{5/2}}\left[
\log{\mu\over4\pi T}-{2\over3}\log\alpha+C_5\right]
-\nonumber\\ &&\nonumber\hphantom{\times\bigg\{}-{15\over16}{\alpha^3}
\times \nonumber\\ && \hphantom{\times\bigg\{}\times\left[
\nonumber-48\log{\mu\over4\pi T}
+16{\zeta^{\prime}(-1)\over\zeta(-1)}
-32\gamma_E-84\zeta(3)+8+16\log{2\over3}
+\right.\\ && \nonumber\hphantom{\times\bigg\{\times\bigg[}\left.
+16\log\alpha
+\pi^2\left(2+12\log2-4\log{2\over3}-4\log\alpha
+8\log{\Lambda\over4\pi T}\right)
\right]
+\nonumber\\ &&\nonumber\hphantom{\times\bigg\{}+{225\sqrt{6}\over8}{\alpha^{7/2}}
\times\nonumber\\&&\hphantom{\times\bigg\{}\times
\left[
\log^2{\mu\over4\pi T}+\left({221\over135}+{2\over3}\gamma_E
-{4\over3}\log{2\over3}-{4\over3}{\zeta^{\prime}(-1)\over\zeta(-1)}
-{4\over3}\log\alpha\right)\log{\mu\over4\pi T}+
\right.\nonumber\\&& \hphantom{\times\bigg\{\times\bigg[}\left. +\left({2\over15}+
{8\over45}{\zeta^{\prime}(-1)\over\zeta(-1)}-{52\over45}\gamma_E
+{8\over45}\log{2\over3}\right)\log\alpha+{4\over45}\log^2\alpha
+\right.\nonumber\\&&\hphantom{\times\bigg\{\times\bigg[}
+C_7\bigg]
\bigg\}\;,
\label{g7soft}
\eqa
where the constants $C_5$ and $C_7$ are defined below.

\section{Results and discussion}
The full pressure is given by minus the sum of Eq.~(\ref{fhard})
and Eq.~(\ref{s15}). The strict weak-coupling result for the pressure
through order $g^7$ is minus the sum of Eq.~(\ref{fhard})
and Eq.~(\ref{g7soft}).
This yields
\begin{eqnarray}
    \mathcal P_{} &=& \mathcal P_\mathrm{ideal} \times\nonumber\\ &&\times\bigg\{
        1
        - \frac{5}{4} \alpha
        + \frac{5 \sqrt 6}{3} \alpha^{3/2}
        + \frac{15}{4} \alpha^2\bigg[
            \log\frac{\mu}{4 \pi T} + C_4
            \bigg]
    -\nonumber \\ \nonumber
    && \hphantom{\times\bigg\{}           
        - \frac{15 \sqrt 6}{2} \alpha^{5/2}\bigg[
            \log\frac{\mu}{4 \pi T} - \frac{2}{3} \log \alpha + C_5
            \bigg] 
    -\nonumber \\ \nonumber
    && \hphantom{\times\bigg\{}
        - \frac{45}{4} \alpha^3\bigg[
            \log^2 \frac{\mu}{4 \pi T}
            - \frac{1}{3} \left(
                \frac{269}{45} - 2 \gamma_E
                - 8 \frac{\zeta'(-1)}{\zeta(-1)}
                + 4 \frac{\zeta'(-3)}{\zeta(-3)}
                \right) \log\frac{\mu}{4 \pi T}
+\nonumber \\ \nonumber
    && \hphantom{\times\bigg\{- \frac{45}{4} \alpha^3\bigg[}                
            + \frac{1}{3} (4 - \pi^2) \log\alpha + C_6
            \bigg] 
+\\ && \nonumber \hphantom{\times\bigg\{}
+{225\sqrt{6}\over8}\alpha^{7/2}\times\nonumber\\ && \hphantom{\times\bigg\{}\times
\left[
\log^2{\mu\over4\pi T}
+\left({221\over135}
+{2\over3}\gamma_E
-{4\over3}\log{2\over3}
-{4\over3}{\zeta^{\prime}(-1)\over\zeta(-1)}
-{4\over3}\log\alpha
\right)
\log{\mu\over4\pi T}
\right.+\nonumber\\ &&\left.
\hphantom{\times\bigg\{\times\bigg[}+\left({2\over15}+
{8\over45}{\zeta^{\prime}(-1)\over\zeta(-1)}
-{52\over45}\gamma_E
-{8\over45}\log{2\over3}\right)\log\alpha
+{4\over45}\log^2\alpha
+C_7
\right]
        \bigg\},
\label{weak66}
\end{eqnarray}
where ${\cal P}_{\rm ideal}=\pi^2T^4/90$ and
where the constants $C_4-C_7$ are 
\begin{eqnarray}
    C_4 &\equiv&
        - \frac{59}{45} + \frac{1}{3} \gamma_E
        + \frac{4}{3} \frac{\zeta'(-1)}{\zeta(-1)}
        - \frac{2}{3} \frac{\zeta'(-3)}{\zeta(-3)}\;,
        \\
    C_5 &\equiv&
        \frac{5}{6}
        + \frac{1}{3} \gamma_E - \frac{2}{3} \log \frac{2}{3}
        - \frac{2}{3} \frac{\zeta'(-1)}{\zeta(-1)}\;,
        \\
    C_6 &\equiv&
        \frac{1}{3} (4 - \pi^2) \log\frac{2}{3}
        + \frac{103}{54} + \frac{1}{18} C'_\mathrm{ball} - \frac{1}{6} C_\mathrm{triangle}^a - \frac{\pi^2}{12} C_\mathrm{triangle}^b+ \frac{4}{9} \gamma_1 - \frac{511}{180} \gamma_E 
       +\nonumber \\
        &&
        + \frac{25}{36} \gamma_E^2
       + \frac{5 \pi^2}{24} - \frac{\pi^2}{3} \gamma_E
        + \pi^2 \log 2
        + \left(\frac{175}{54} - \frac{1}{9} \gamma_E\right) \frac{\zeta'(-1)}{\zeta(-1)}
        + \frac{2}{3} \left(\frac{\zeta'(-1)}{\zeta(-1)}\right)^2
                +\nonumber \\
        &&        
        + \frac{5}{9} \frac{\zeta''(-1)}{\zeta(-1)}
        - \frac{2}{3} \gamma_E \frac{\zeta'(-3)}{\zeta(-3)}
-{2267\over324}\zeta(3)\;, \\ \nonumber
C_7&=&-{1457\over810}
+{1\over45}C^{\prime}_{\rm ball}
-{2\over15}C_I+{749\over270}\gamma_E+{56\over15}\gamma_1
-{11\over20}\pi^2+
{2\over15}\frac{\zeta^{\prime\prime}(-1)}{\zeta(-1)}
+\\ &&\nonumber
+{16\over15}\gamma_E\log(2\pi)-{16\over15}\log^2(2\pi)
-{52\over45}\gamma_E\log{2\over3}
-{19\over27}\frac{\zeta^{\prime}(-1)}{\zeta(-1)}
-{38\over45}\gamma_E\frac{\zeta^{\prime}(-1)}{\zeta(-1)}
+\\&&\nonumber
+{4\over45}\left(\frac{\zeta^{\prime}(-1)}{\zeta(-1)}\right)^2
+{34\over15}\log{2\over3}
+{2\over5}\pi^2\log2
+{4\over45}\log^23
+{28\over15}\log^22
-\\&&
-{8\over45}\log2\log3
+{8\over45}\frac{\zeta^{\prime}(-1)}{\zeta(-1)}\log{2\over3}
-{97\over54}\zeta(3)+{16\over9}{\rm Li}_2(\mbox{$1\over4$})
+{97\over90}\gamma_E^2
+{16\over5}C_{\rm 4j}\;,
\end{eqnarray}
where $C_{\rm triangle}^a=-25.7055$ and $C_{\rm triangle}^b=28.9250$.
The numerical values of $C_4-C_7$ are 
\begin{eqnarray}
    C_4 &=& 1.09775\;, \\
    C_5 &=& -0.0273205\;, \\
    C_6 &=& -6.5936\;, \\
    C_7 &=&  -0.862\;.
\end{eqnarray}
Note that the $\Lambda$-dependence cancels in the result~(\ref{weak66}).
Using Eq.~(\ref{arun})
for the running of $\alpha$, 
it is straightforward to check that the final result
Eq.~(\ref{weak66}) is RG invariant up to higher-order corrections.
The order-$g^4$ result was obtained by Frenkel, Saa, and 
Taylor~\cite{Frenkel:1992az}, the order-$g^5$ result by Parwani and 
Singh~\cite{Parwani:1994zz}, 
the order-$g^6\log(g)$ result by Braaten and Nieto~\cite{Braaten:1995cm},
and the order-$g^6$ result by
Gynther {\it et al}~\cite{Gynther:2007bw}. The latter was later 
reproduced in Ref.~\cite{Andersen:2008bz} using screened perturbation 
theory~\cite{Karsch:1997gj,Andersen:2000yj,Andersen:2001ez}
by taking the weak-coupling limit for the
mass parameter, $m=gT/\sqrt{24}$.

An expansion of the pressure in powers of $g$ is given in Eq.~(\ref{weak66}).
It is accurate up to corrections of order $g^8\log(g)$.
A more accurate expression
can be obtained by using the fact that our short-distance
coefficients satisfy a set of evolution equations.
The solutions to the evolution equations are
\bqa
\label{rg1}
g_3^2(\Lambda)&=&g_3^2(2\pi T)\;,\\ 
f(\Lambda)&=&f(2\pi T)-{\pi^2g_3^6(2\pi T)\over192(4\pi)^4}
\log{\Lambda\over2\pi T}
\label{rgmid}
\;,\\ 
m^2(\Lambda)
&=&m^2(2\pi T)+{g_3^4(2\pi T)\over6(4\pi)^2}\log{\Lambda\over2\pi T}\;.
\label{rg2}
\eqa
If we substitute the short-distance 
coefficients~(\ref{rg1}) and (\ref{rg2}) into Eq.~(\ref{s15}) 
and add the short-distance contribution~(\ref{rgmid}),  
setting $\Lambda=gT/\sqrt{24}$ everwhere, and expand the resulting expression  
in powers of $g$, 
we obtain 
the complete result for the pressure,
which is correct up to order $g^8\log (g)$.
The contributions to the free energy ${\cal F}$
of order $g^8\log(g)$ come from~(\ref{rgmid}) and from using~(\ref{rg2}) to expand the $g_3^2m^2T$ term in~(\ref{s15}). This yields

\bqa
{\cal F}_{\rm g^8\log(g)}&=&
{3g^8T^4\over64(4\pi)^6}\left(
\log2-\gamma_E
\right)(4-\pi^2)\log(g)\;. 
\label{g8log}
\eqa
Moreover, using the solutions to the flow equations, we are summing
up leading logarithms of the form $g^{2n+3}\log^n(g)$ 
and e.g. 
subleading logarithms of the form $g^{2n+5}\log^n(g)$, where $n=2,3,...$.
These terms are obtained by expanding out
the $m^3T$ and $g_3^4mT$ terms in~(\ref{s15}), respectively.



\DOUBLEFIGURE{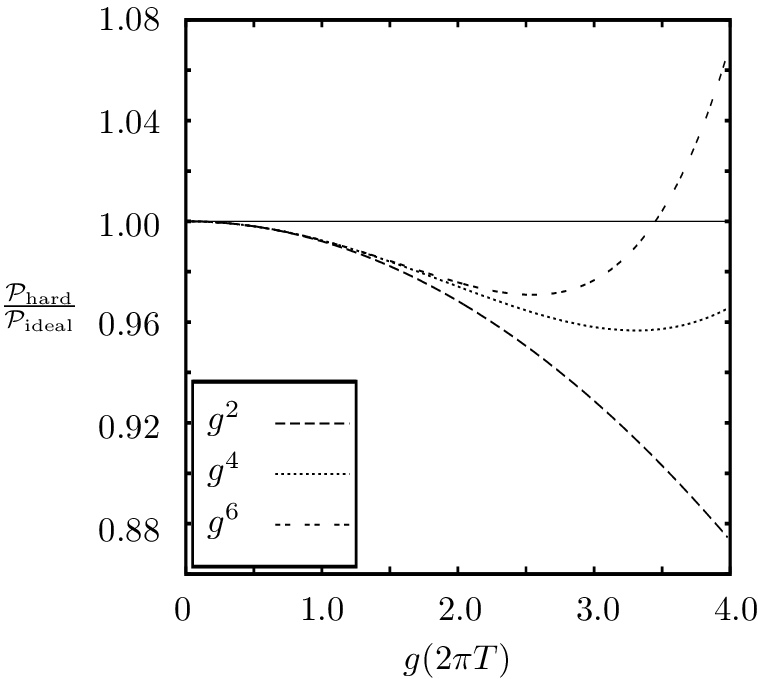,width=6.8cm}{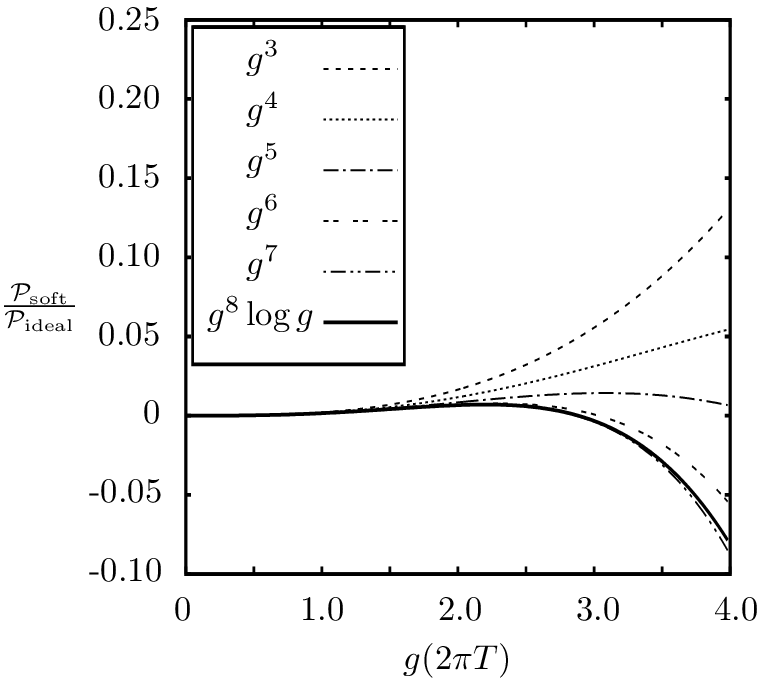,width=6.8cm}{Hard contributions ${\cal P}_{\rm hard}$
to the pressure ${\cal P}$ normalized to ${\cal P}_{\rm ideal}$ to order $g^2$,
$g^4$, and $g^6$.\label{g1g7}}{Soft contributions ${\cal P}_{\rm soft}$
to the pressure ${\cal P}$ normalized to ${\cal P}_{\rm ideal}$ to order $g^3$,
$g^4$, $g^5$, $g^6$, $g^7$, and $g^8\log(g)$.\label{sptweak}}

In Fig.~\ref{g1g7}, we show the various loop orders of
${\cal P}_{\rm hard}$ normalized to 
${\cal P}_{\rm ideal}$ to orders $g^2$, $g^4$, and $g^6$,
where ${\cal P}_{\rm hard}$ is given by minus 
Eq.~(\ref{fhard})~\footnote{Note that we omit the pole in 
$\epsilon$ in Eq.~(\ref{fhard}) in the plots of the hard part.}.
We have chosen $\mu=2\pi T$ and $\Lambda=2\pi T$.
We notice that the successive approximations are larger than the
previous one.
In Fig.~\ref{sptweak}, we show the weak-coupling expansion of
${\cal P}_{\rm soft}$ normalized to ${\cal P}_{\rm ideal}$ to orders $g^3$, 
$g^4$, $g^5$ $g^6$, $g^7$, and $g^8\log(g)$, 
where ${\cal P}_{\rm soft}$ is given by minus the sum of Eqs.~(\ref{g7soft})
and~(\ref{g8log}).

\DOUBLEFIGURE{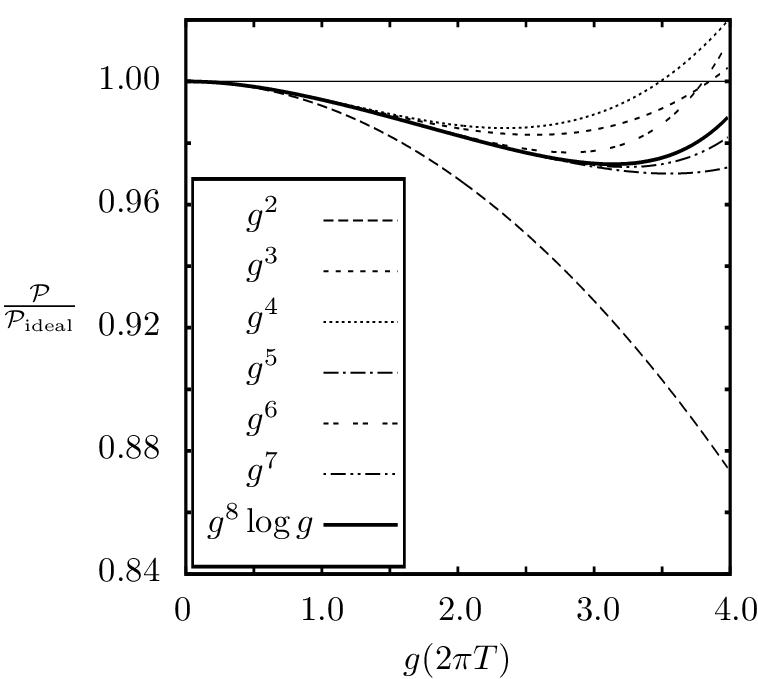,width=6.8cm}{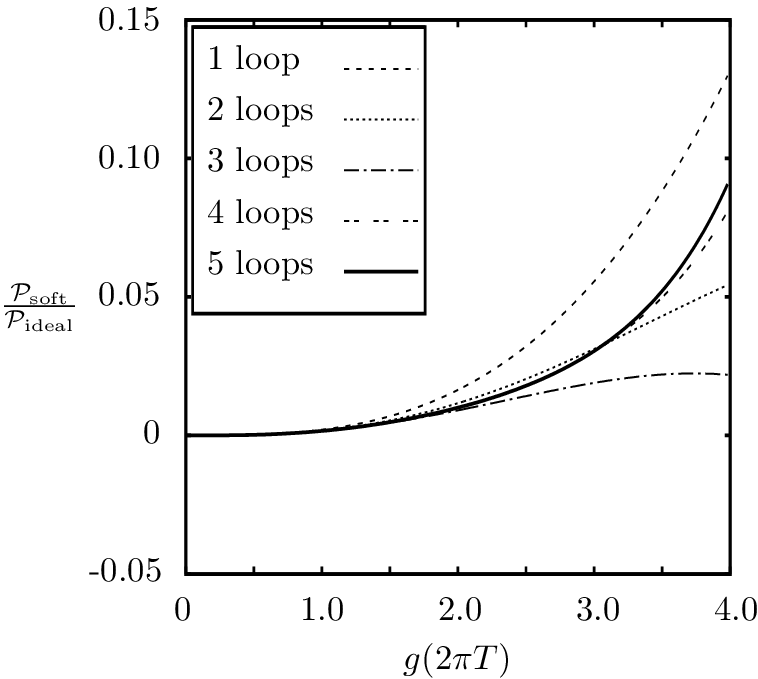,width=6.8cm}{Weak-coupling expansion of the pressure ${\cal P}$
normalized to ${\cal P}_{\rm ideal}$ to order $g^2$, $g^3$,
$g^4$, $g^5$, $g^6$, $g^7$, and $g^8\log(g)$.\label{totalweak}}{Soft contributions ${\cal P}_{\rm soft}$
to the pressure ${\cal P}$
normalized to ${\cal P}_{\rm ideal}$ at one through five loops.\label{selsoft}}

In Fig.~\ref{totalweak}, we show the weak-coupling expansion of
the pressure 
${\cal P}_{}$ given by ~(\ref{weak66}) minus ~(\ref{g8log})
normalized to ${\cal P}_{\rm ideal}$ to orders $g^2$, $g^3$ 
$g^4$, $g^5$ $g^6$, $g^7$, and $g^8\log(g)$ . 
The convergence properties of the 
successive approximations of the sum 
${\cal P}={\cal P}_{\rm hard}+{\cal P}_{\rm soft}$ clearly is
better than the convergence properties of 
the successive approximations to 
${\cal P}_{\rm hard}$ and ${\cal P}_{\rm soft}$ separately.

In Fig.~\ref{selsoft}, we plot the successive loop orders of minus
Eq.~(\ref{s15}) normalized to ${\cal P}_{\rm ideal}$.
In the one- and two-loop approximations, we use the
leading-order results for $g_3^2$ and for $m^2$. At three and four loops,
we use the leading-order result for $g_3^2$ and next-to-leading order
result for $m^2$. 
Finally, at five loops, we use the solutions to the 
evolution equations for $g_3^2$, $f$, and $m^2$.
The renormalization scale is $\mu=2\pi T$ and the factorization
scale is $\Lambda=gT/\sqrt{24}$.
These approximations represent a selective resummation of higher-order terms.
Clearly, the convergence is better than the strict perturbative
expansion. In particular, the three-, four-, and five-loop approximations
are very close.

\EPSFIGURE{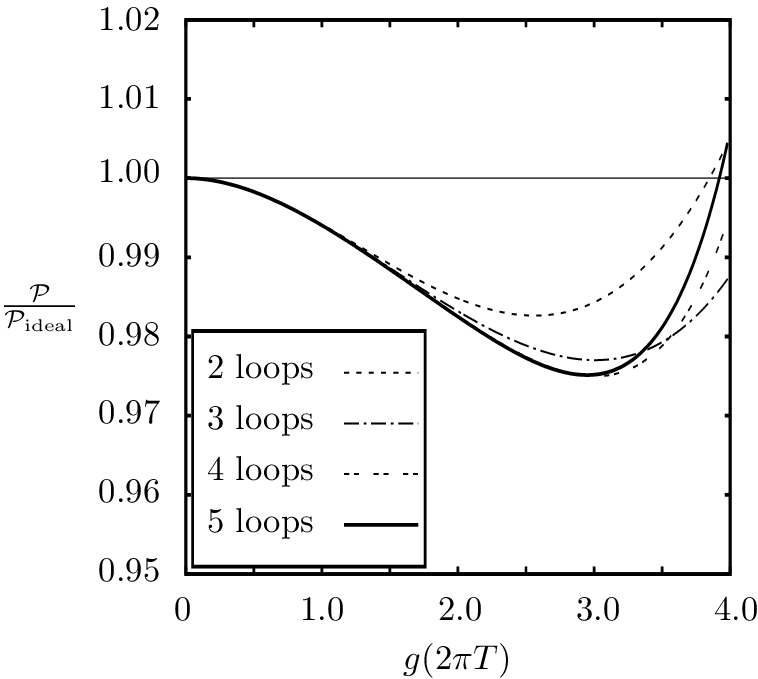,width=6.8cm}{Successive approximations to the pressure ${\cal P}$
normalized to ${\cal P}_{\rm ideal}$ at two through five loops.,\label{seltotal}}
In Fig.~\ref{seltotal}, we plot the successive loop orders of the
the pressure which is given by the 
sum of minus Eq.~(\ref{fhard}) minus
Eq.~(\ref{s15}), and minus~(\ref{g8log}),
normalized to ${\cal P}_{\rm ideal}$, starting at two loops.
We are using the
same approximations for $g_3^2$ and $m^2$ as in the previous plot.
Again we notice that the convergence of $P_{}$ is better than 
$P_{\rm hard}$ and $P_{\rm soft}$ separately. In fact the convergence
is very good as the 3-loop through 5-loop approximations are very close.
It is not surprising that a selective resummation improves the convergence
of the series. This was also notice in screened perturbation 
theory~\cite{Karsch:1997gj,Andersen:2000yj,Andersen:2001ez,Andersen:2008bz}.


\section{Summary}
In the present paper, we have calculated the pressure to order
$g^8\log(g)$ in massless 
$\phi^4$-theory at weak coupling. 
The first step is the determination of the coefficients in the 
dimensionally reduced effective field theory. This calculation encodes
the physics of the hard scale $T$. The mass parameter was needed to
order $g^6$ and involves a nontrivial three-loop sum-integral that
was recently calculated in Ref.~\cite{Andersen:2008bz}.
The second step consists of using the effective theory to calculate
the vacuum diagrams through five loops.
All loop diagrams in the effective theory but one could be calculated
analytically with dimensional regularization.
This way of organizing the calculations is more economical and efficient
than resummed perturbation theory.

The parameters of the effective theory, $g_3^2$, $f$, and $m^2$,
satisfy a set of evolution equations. The solutions of these equations
show that the parameters depend explicitly on the renormalization scale.
This dependence is necessary to cancel the dependence on the scale
in the effective theory~\cite{Braaten:1995cm}.
The fact the our final result for the pressure is independent of 
the renormalization scale is a nontrivial check of the calculations.
Furthermore, by choosing $\Lambda=gT/\sqrt{24}$
and using the solutions to the evolution equations, we were able to 
sum up leading logarithms of the form $g^{2n+3}\log^n(g)$ 
and e.g. 
subleading logarithms of the form $g^{2n+5}\log^n(g)$, where $n=2,3,...$.
as well as obtaining the coefficient of the $g^8\log(g)$ term.


As pointed out in Ref.~\cite{Gynther:2007bw}, it would be advantageous
to develop the machinery of calculating complicated
multiloop sum-integrals in an automated fashion as has been done
for Feynman diagrams at zero temperature. Perhaps such techniques
could provide analytical expressions for the constants that today are known
only numerically. This is necessary if one wants to tackle the 
formidable problem of calculating the hard part of the $g^6$-contribution
to the free energy of QCD.

\section*{Acknowledgments}
The authors would like to thank T. Brauner, B. Kastening,
and M. Laine for useful discussions.
The authors would like to thank the
Niels Bohr International Academy for kind hospitality.

\appendix
\renewcommand{\theequation}{\thesection.\arabic{equation}}
\section{Sum-integrals}\label{appa}
In the imaginary-time formalism for thermal field theory, 
the 4-momentum $P=(P_0,{\bf p})$ is Euclidean with $P^2=P_0^2+{\bf p}^2$. 
The Euclidean energy $p_0$ has discrete values:
$P_0=2n\pi T$ for bosons,
where $n$ is an integer. 
Loop diagrams involve sums over $P_0$ and integrals over ${\bf p}$. 
With dimensional regularization, the integral is generalized
to $d = 3-2 \epsilon$ spatial dimensions.
We define the dimensionally regularized sum-integral by
\bqa
  \hbox{$\sum$}\!\!\!\!\!\!\int_{P}& \;\equiv\; &
  \left(\frac{e^\gamma\mu^2}{4\pi}\right)^\epsilon\;
  T\!\!\!\!\!\!\sum_{P_0=2n\pi T}\:\int {d^{3-2\epsilon}p \over (2 \pi)^{3-2\epsilon}}\;,
\label{sumint-def}
\eqa
where $3-2\epsilon$ is the dimension of space and $\mu$ is an arbitrary
momentum scale. 
The factor $(e^\gamma/4\pi)^\epsilon$
is introduced so that, after minimal subtraction 
of the poles in $\epsilon$
due to ultraviolet divergences, $\mu$ coincides 
with the renormalization
scale of the $\overline{\rm MS}$ renormalization scheme.

\subsection{One-loop sum-integrals}
The massless one-loop sum-integral is given by 
\bqa\nonumber
{\cal I}_n&\equiv&
\sumint_{P}{1\over P^{2n}}\\
&=&(e^{\gamma_E}\mu^2)^{\epsilon}{\zeta(2n-3+2\epsilon)\over8\pi^2}
{\Gamma(n-\mbox{$3\over2$}+\epsilon)
\over\Gamma(\mbox{$1\over2$})\Gamma(n)} 
(2\pi T)^{4-2n-2\epsilon}\;,
\eqa
where $\zeta(x)$ is Riemann's zeta function.
Specifically, we need the sum-integrals
\bqa\nonumber
{\cal I}_0^{\prime}&\equiv&
\sumint_P\log P^2
\\ &=&
-{\pi^2T^4\over45}\left[1+{\cal O}\left(\epsilon\right)\right]\;,\\
{\cal I}_1
&=&{T^2\over12}\left({\mu\over4\pi T}\right)^{2\epsilon}
\left[1+\left(2+2{\zeta^{\prime}(-1)\over\zeta(-1)}\right)\epsilon
+\right.\nonumber\\ && \hphantom{{T^2\over12\bigg[}\left({\mu\over4\pi T}\right)^{2\epsilon}}\left.
+\left(4+{\pi^2\over4}
+4{\zeta^{\prime}(-1)\over\zeta(-1)}
+2{\zeta^{\prime\prime}(-1)\over\zeta(-1)}
\right)\epsilon^2
+{\cal O}\left(\epsilon^3\right)\right]\;,\\
{\cal I}_2&=&
{1\over(4\pi)^2}\left({\mu\over4\pi T}\right)^{2\epsilon}
\left[{1\over\epsilon}
+2\gamma_E+\left({\pi^2\over4}-4\gamma_1\right)\epsilon\
+{\cal O}\left(\epsilon^2\right)\right]\;,\\
{\cal I}_3
&=&{1\over(4\pi)^4T^2}\left[2\zeta(3)+
{\cal O}\left(\epsilon\right)\right]\;.
\eqa

\subsection{Two-loop sum-integrals}
We need three two-loop sum-integral that are listed below:
\bqa\nonumber
{\cal I}_{\rm sun}&=&
\sumint_{PQ}{1\over P^2Q^2(P+Q)^2} \\
&=&{\cal O}(\epsilon)\;,\\
\sumint_{PQ}{P^2+(2/d)p^2\over P^6Q^2(P+Q)^2}&=&
{3\over4(4\pi)^4}\left({\mu\over4\pi T}\right)^{4\epsilon}
\left[{1\over\epsilon^2}+\left({5\over6}+4\gamma_E\right)
{1\over\epsilon}
+{89\over36}+{\pi\over2}
+\right.\nonumber\\ && \hphantom{{3\over4(4\pi)^4}\left({\mu\over4\pi T}\right)^{4\epsilon}\bigg[}\left.
+{10\over3}\gamma_E+4\gamma_E^2-8\gamma_1+{\cal O}(\epsilon)
\right]\;,\\
\sumint_{PQ}{P^2-(4/d)p^2\over P^6Q^2(P+Q)^2}&=&
{1\over4(4\pi)^4}\left({\mu\over4\pi T}\right)^{4\epsilon}
\left[
{1\over\epsilon}+{19\over6}+4\gamma_E+{\cal O}(\epsilon)
\right]\;.
\eqa
The setting-sun sum-integral was first calculated by Arnold and Zhai in
Refs.~\cite{Arnold:1994ps,Arnold:1994eb}. The remaining two-loop sum-integrals were
calculated by Braaten and Petitgirard~\cite{Braaten:2001en,Braaten:2001vr} using the
techniques developed in~\cite{Arnold:1994ps,Arnold:1994eb}.

\subsection{Three-loop sum-integrals}
We need the following three-loop sum-integrals:
\bqa\nonumber
{\cal I}_{\rm ball}
&=&\sumint_{PQR}{1\over P^2Q^2R^2(P+Q+R)^2}\\
&=&{T^4\over24(4\pi)^2}\left({\mu\over4\pi T}\right)^{6\epsilon}\left[
{1\over\epsilon}+{91\over15}+8{\zeta^{\prime}(-1)\over\zeta(-1)}
-2{\zeta^{\prime}(-3)\over\zeta(-3)}
+{\cal O}(\epsilon)
\right]\;, 
\\ \nonumber
{\cal I}_{\rm ball}^{\prime}
&=&
\sumint_{PQR}{1\over P^4Q^2R^2(P+Q+R)^2} \\ \nonumber
&=&{T^2\over8(4\pi)^4}\left({\mu\over4\pi T}\right)^{6\epsilon}
\left[
{1\over\epsilon^2}+
\left({17\over6}+4\gamma_E
+2{\zeta^{\prime}(-1)\over\zeta(-1)}\right)
{1\over\epsilon}+
\right.\\&&\left.
+{1\over2}
\gamma_E\left(17+15\gamma_E+12{\zeta^{\prime}(-1)\over\zeta(-1)}\right)
+C_{\rm ball}^{\prime}
+{\cal O}(\epsilon)
\right]\;,
\label{bp222}
\eqa
and
\bqa 
\sumint_P{1\over P^2}&&\hspace{-0.5cm} \left\{[\Pi(P)]^2-{2\over(4\pi)^2\epsilon}
\Pi(P)\right\}
    =\nonumber\\ 
      &=&  -\frac{T^2}{4(4\pi)^4}\left({\mu\over4\pi T}\right)^{6\epsilon}
      \times \nonumber\\ && \times
 \bigg\{
            \frac{1}{\epsilon^2}
            + \frac{1}{\epsilon} \left[
\frac{4}{3}+2\frac{\zeta'(-1)}{\zeta(-1)} + 4\gamma_E 
                \right]
    +\nonumber \\   && \nonumber
\hphantom{\times\bigg\{}            + \frac{1}{3} \bigg[46 - 8\gamma_E 
- 16\gamma_E^2 
- 104\gamma_1 - 24\gamma_E \log(2\pi)
                + 24\log^2(2\pi) + 
\frac{45\pi^2}{4} 
+\\  &&
\hphantom{\times\bigg\{+{1\over3}\bigg[}+ 24\frac{\zeta'(-1)}{\zeta(-1)}
+ 2\frac{\zeta''(-1)}{\zeta(-1)}
                + 16\gamma_E \frac{\zeta'(-1)}{\zeta(-1)}
                \bigg]
+C_I
+{\cal O}(\epsilon)            \bigg\}\;,
\label{oss}
\eqa
where the self-energy $\Pi(P)$ is defined as
\bqa
{\Pi}(P)&=&\sumint_Q{1\over Q^2(P+Q)^2}\;,
\label{pidef1}
\eqa
and $C_{\rm ball}^{\prime}=48.7976$ and $C_I=-38.4672$.
The massless basketball sum-integral was first calculated by Arnold and Zhai
in Refs.~\cite{Arnold:1994ps,Arnold:1994eb}. The sum-integral~Eq.~(\ref{bp222}) 
was calculated by Gynther {\it et al.} in Ref.~\cite{Gynther:2007bw}.
The expression for the 
sum-integral Eq.~(\ref{oss}) was calculated in Ref.~\cite{Andersen:2008bz}.

\subsection{Four-loop sum-integrals}
We also need a single four-loop sum-integral which was calculated in
Ref.~\cite{Gynther:2007bw}:
\bqa\nonumber
\sumint_{P}&&\hspace{-0.5cm}\left\{[\Pi(P)]^3-{3\over(4\pi)^2\epsilon}[\Pi(P)]^2\right\}
=\nonumber\\
&=&-{T^4\over16(4\pi)^4}\times\nonumber\\ &&\times
\left[{1\over\epsilon^2}+\left(4\log{\mu\over4\pi T}+
{10\over3}+4{\zeta^{\prime}(-1)\over\zeta(-1)}
\right){1\over\epsilon}
+(2\log{\mu\over4\pi T}+\gamma_E)^2
+\right.\nonumber\\ &&\left.\nonumber
\hphantom{\times\bigg[}+\left({6\over5}-2\gamma_E
+4{\zeta^{\prime}(-3)\over\zeta(-3)}\right)(2\log{\mu\over4\pi T}+\gamma_E)
+C_{\rm triangle}^a
\right]
-\nonumber\\ &&
-{T^4\over512(4\pi)^2}\left[
{1\over\epsilon}+8\log{\mu\over4\pi T}
+4\gamma_E
+C_{\rm triangle}^b
\right]
+{\cal O}(\epsilon)\;,
\eqa
where $C_{\rm triangle}^a=-25.7055$ and $C_{\rm triangle}^b=28.9250$.

\section{Three-dimensional integrals}\label{appb}
Dimensional regularization can be used to
regularize both the ultraviolet divergences and infrared divergences
in 3-dimensional integrals over momenta.
The spatial dimension is generalized to  $d = 3-2\epsilon$ dimensions.
Integrals are evaluated at a value of $d$ for which they converge and then
analytically continued to $d=3$.
We use the integration measure
\begin{equation}
 \int_p\;\equiv\;
  \left(\frac{e^\gamma\mu^2}{4\pi}\right)^\epsilon\;
\:\int {d^{3-2\epsilon}p \over (2 \pi)^{3-2\epsilon}}\;.
\label{int-def}
\end{equation}

\subsection{One-loop integrals}
The one-loop integral is given by
\bqa\nonumber
I_n&\equiv&\int_p{1\over(p^2+m^2)^n}\\
&=&{1\over8\pi}(e^{\gamma_E}\mu^2)^{\epsilon}
{\Gamma(n-\mbox{$3\over2$}+\epsilon)
\over\Gamma(\mbox{$1\over2$})
\Gamma(n)}m^{3-2n-2\epsilon}
\;.
\eqa
Specifically, we need
\bqa\nonumber
I_0^{\prime}&\equiv&
\int_p\log(p^2+m^2)\\
&=&
-{m^3\over6\pi}\left({\mu\over2m}\right)^{2\epsilon}
\left[
1
+{8\over3}
\epsilon
+\left(
{52\over9}+{\pi^2\over4}\right)\epsilon^2
+{\cal O}
\left(\epsilon^3\right)
\right]\;,\\ 
I_1&=&-{m\over4\pi}\left({\mu\over2m}\right)^{2\epsilon}
\left[
1+2
\epsilon+\left(
4+{\pi^2\over4}\right)\epsilon^2
+{\cal O}\left(\epsilon^3\right)
\right]\;,\\
\label{i2}
I_2&=&{1\over8\pi m}\left({\mu\over2m}\right)^{2\epsilon}
\left[
1
+{\pi^2\over4}
\epsilon^2
+{\cal O}\left(\epsilon^3\right)
\right]\;,\\
I_3&=&{1\over32\pi m^3}\left({\mu\over2m}\right)^{2\epsilon}
\left[
1+2
\epsilon
+{\pi^2\over4}\epsilon^2
+{\cal O}\left(\epsilon^3\right)
\right]\;,
\\ 
I_4&=&{1\over64\pi m^5}\left({\mu\over2m}\right)^{2\epsilon}
\left[
1+{8\over3}
\epsilon
+\left({4\over3}+{\pi^2\over4}\right)\epsilon^2
+{\cal O}\left(\epsilon^3\right)
\right]
\;.
\eqa
\subsection{Two-loop integrals}
We need the following two-loop integral
\bqa\nonumber
I_{\rm sun}(p=im)&=&
\int_{qr}{1\over q^2+m^2}{1\over r^2+m^2}
{1\over({\bf p}+{\bf q}+{\bf r})^2+m^2}\bigg|_{p=im} \\
&=&
{1\over4(4\pi)^2}
\left({\mu\over2m}\right)^{4\epsilon}\times\nonumber\\&&\times
\left[
{1\over\epsilon}+6-8\log2+\left(36-{\pi^2\over6}-48\log2+8\log^22\right)
\epsilon
+{\cal O}(\epsilon^2)
\right].
\label{isunp}
\eqa
This integral was calculated to order $\epsilon^0$ in Ref.~\cite{Braaten:1995cm}
and to order $\epsilon$ in Refs.~\cite{Braaten:2001en,Braaten:2001vr}.

\subsection{Three-loop integrals}
We need the following three-loop integrals:
\bqa\nonumber
I_{\rm ball}&=&
\int_{pqr}{1\over p^2+m^2}{1\over q^2+m^2}{1\over r^2+m^2}
{1\over({\bf p}+{\bf q}+{\bf r})^2+m^2}
\\ &=&
-{m\over(4\pi)^3}\left({\mu\over2m}\right)^{6\epsilon}
\times\nonumber\\&&\times
\left[{1\over\epsilon}+8-4\log2
+4\left(13+{17\over48}\pi^2-8\log2+\log^22\right)\epsilon
+{\cal O}\left(\epsilon^2\right)
\right]\;,
\\ \nonumber
I_{\rm ball}^{\prime}&=&
\int_{pqr}{1\over(p^2+m^2)^2}{1\over q^2+m^2}{1\over r^2+m^2}
{1\over({\bf p}+{\bf q}+{\bf r})^2+m^2} \\
&=&
{1\over8m(4\pi)^3}\left({\mu\over2m}\right)^{6\epsilon}
\times\nonumber\\&&\times
\left[{1\over\epsilon}+2-4\log2
+4\left(1+{17\over48}\pi^2-2\log2+\log^22\right)\epsilon
+{\cal O}\left(\epsilon^2\right)\right]\;,
\label{iballp}
\\ \nonumber
J&=&
\int_{pqr}{1\over(q^2+m^2)^2}
{1\over({\bf p}+{\bf q})^2+m^2}
{1\over(r^2+m^2)^2}
{1\over({\bf p}+{\bf r})^2+m^2}\\
&=&{1\over16m^3(4\pi)^3}\left({\mu\over2m}\right)^{6\epsilon}
\left[1+{\cal O}\left(\epsilon\right)\right]\;,
\label{jdef}
\\ \nonumber
K&=&
\int_{pqr}{1\over(q^2+m^2)^3}
{1\over({\bf p}+{\bf q})^2+m^2}
{1\over r^2+m^2}
{1\over({\bf p}+{\bf r})^2+m^2}\\
&=&{1\over32m^3(4\pi)^3}
\left({\mu\over2m}\right)^{6\epsilon}\left[{1\over\epsilon}+5-4\log2
+{\cal O}\left(\epsilon\right)\right]\;.
\eqa
The massive basketball was calculated
in Ref.~\cite{Braaten:1995cm} to order $\epsilon^0$, and to order
$\epsilon$ in Ref.~\cite{Kajantie:2003ax}. 
$I_{\rm ball}^{\prime}$ can be obtained by differentiation of
$I_{\rm ball}$ with respect to $m$.
The 3-loop integrals $J$ and $K$ are calculated in Appendix C.
\subsection{Four-loop integrals}
We need the following two four-loop integrals
\bqa\nonumber
I_{\rm triangle}&=&
\int_{pqrs}{1\over q^2+m^2}{1\over({\bf p}+{\bf q})^2+m^2}
{1\over r^2+m^2}{1\over({\bf p}+{\bf r})^2+m^2}
{1\over s^2+m^2}{1\over({\bf p}+{\bf s})^2+m^2}
\\
&=&{\pi^2\over32(4\pi)^4}\left({\mu\over2m}\right)^{8\epsilon}\left[
{1\over\epsilon}+2+4\log2
-{84\over\pi^2}\zeta(3)+{\cal O}\left(\epsilon\right)
\right]\;,\\\nonumber
I_{\rm triangle}^{\prime}&=&
\int_{pqrs}{1\over(q^2+m^2)^2}{1\over({\bf p}+{\bf q})^2+m^2}
{1\over r^2+m^2}{1\over({\bf p}+{\bf r})^2+m^2}
{1\over s^2+m^2}{1\over({\bf p}+{\bf s})^2+m^2}
\\
&=&{\pi^2\over48m^2(4\pi)^4}\left({\mu\over2m}\right)^{8\epsilon}\left[
1
+{\cal O}\left(\epsilon\right)
\right]\;. 
\eqa
The triangle diagram was calculated in Ref.~\cite{Vuorinen:2004rd}. 
The diagram $I_{\rm triangle}^{\prime}$ follows from the triangle diagram
upon differentiation with respect to $m^2$.

\subsection{Five-loop integrals}
\bqa\nonumber
I_{\rm rung}
&=&
\int_{pqrst}{1\over q^2+m^2}{1\over({\bf p}+{\bf q})^2+m^2}
{1\over r^2+m^2}{1\over({\bf p}+{\bf r})^2+m^2}
{1\over s^2+m^2}{1\over({\bf p}+{\bf s})^2+m^2}
\times \\ \nonumber
&&\hphantom{\int_{pqrst}}\times {1\over t^2+m^2}{1\over({\bf p}+{\bf t})^2+m^2} 
\\
&=&
{1\over2m(4\pi)^5}\left({\mu\over2m}\right)^{10\epsilon}\left[
\pi^2\log2-{9\over2}\zeta(3)
+{\cal O}\left(\epsilon\right)
\right]\;, 
\label{irung}\\ \nonumber
I_{\rm double sun}&=&
\int_{pqrst}{1\over(p^2+m^2)^2}{1\over q^2+m^2}{1\over r^2+m^2}
{1\over({\bf p}+{\bf q}+{\bf r})^2+m^2}
\times\nonumber\\ && \hphantom{\int_{pqrst}} \times
{1\over s^2+m^2}{1\over t^2+m^2}{1\over({\bf p}+{\bf s}+{\bf t})^2+m^2}\nonumber\\
\nonumber
&=&
{1\over32m(4\pi)^5}\left({\mu\over2m}\right)^{10\epsilon}
\times\nonumber\\ && \times
\bigg[{1\over\epsilon^2}
+\left(4-8\log2\right){1\over\epsilon}-4+{31\over12}\pi^2
-96\log3
+64\log2+104\log^22
\nonumber\\ &&\hphantom{\times\bigg[}
+80{\rm Li}_2(\mbox{$1\over4$})+{\cal O}\left(\epsilon\right)\bigg]\;,
\\ \nonumber
I_{\rm 4j}
&=&
\int_{pqrst}{1\over q^2+m^2}{1\over({\bf p}+{\bf q})^2+m^2}
{1\over({\bf p}+{\bf r})^2+m^2}{1\over({\bf t}+{\bf r})^2+m^2}
{1\over r^2+m^2}
\times\nonumber\\ && \hphantom{\int_{pqrst}}\times
{1\over({\bf p}+{\bf s})^2+m^2}{1\over({\bf t}+{\bf s})^2+m^2}
{1\over s^2+m^2}
\nonumber\\ &=&
{1\over m(4\pi)^5}\left({\mu\over2m}\right)^{10\epsilon}
\left[C_{\rm 4j}+{\cal O}(\epsilon)\right]
\label{i5c}
\;,
\eqa
where $C_{\rm 4j}=0.443166$.
The integrals are calculated in Appendix C.

\section{Explicit calculations}
In this appendix, we calculate explicitly 
some of the multi-loop vacuum diagrams in three dimensions. 

The three-loop integral $J$ in Eq.~(\ref{jdef}) can be written as
\bqa
\label{J}
J&=&\int_p\left[I_{\rm bubble}^{\prime}(p)\right]^2\;,
\eqa
where
\bqa
I_{\rm bubble}^{\prime}(p)&=&
\int_q{1\over(q^2+m^2)^2}{1\over({\bf p}+{\bf q})^2+m^2}\;.
\eqa
By power counting it is easy to see that both $J$
and $I_{\rm bubble}^{\prime}$ are finite  
in three spatial dimension. The latter then reduces to
\bqa
\label{bubblep}
I_{\rm bubble}^{\prime}(p)&=&{1\over8\pi m}{1\over p^2+4m^2}\;.
\eqa
Inserting Eq.~(\ref{bubblep}) into Eq.~(\ref{J}) and using
Eq.~(\ref{i2}) with $\epsilon=0$ and a mass of 
$2m$. we obtain Eq.~(\ref{jdef}).

The integral $K$ can be calculated by noting the relation
\bqa
I_{\rm ball}^{\prime\prime}&=&-2K-3J\;.
\eqa

The integral $I_{\rm rung}$ in~(\ref{irung}) can be written as 
\bqa
I_{\rm rung}&=&\int_pI_{\rm bubble}^4(p)\;,
\eqa
where
\bqa
I_{\rm bubble}(p)&=&
\int_q{1\over q^2+m^2}{1\over({\bf p}+{\bf q})^2+m^2}\;.
\eqa
The integrals $I_{\rm rung}$ and $I_{\rm bubble}(p)$ 
are convergent in three dimensions. The latter then reduces to
\bqa
I_{\rm bubble}(p)&=&
{1\over4\pi p}\arctan{p\over2m}\;.
\eqa
$I_{\rm rung}$ can now be easily found and the result is
given by Eq.~(\ref{irung}).

The diagram appearing in ${\cal F}_{5i}$
can be written as 
\bqa
\label{dsdef}
I_{\rm double sun}&=&
\int_p{1\over(p^2+m^2)^2}I^2_{\rm sun}(p)\;,
\eqa
where $I_{\rm sun}(p)$ is
\bqa
I_{\rm sun}(p)&=&\int_{qr}
{1\over q^2+m^2}{1\over r^2+m^2}{1\over({\bf p}+{\bf q}+{\bf r})^2+m^2}\;.
\eqa 
In order to isolate the divergences in~(\ref{dsdef}), we 
add and subtract $I_{\rm sun}(p=im)$, and rewrite it as
 \bqa
 I_{\rm double sun}&=&
\int_p{1\over(p^2+m^2)^2}\times\nonumber\\ &&\hphantom{\int_p}\times\left\{\left[I_{\rm sun}(p)-I_{\rm sun}(p=im)
\right]^2+2I_{\rm sun}(p)I_{\rm sun}(p=im)
-\right.\nonumber\\ &&\left.\hphantom{\int_p\times\big\{}
-I^2_{\rm sun}(p=im)
\right\}
\;.
\eqa
We denote the three terms above by $I_{\rm ds1}$, $I_{\rm ds2}$, 
and $I_{\rm ds3}$.
We first consider $I_{\rm ds1}$.
The difference $I_{\rm sun}(p)-I_{\rm sun}(p=im)$ is finite and can be 
calculated directly in three dimensions. 
We obtain
\bqa
I_{\rm sun}&&\hspace{-0.6cm}(p)-I_{\rm sun}(p=im)
=\nonumber\\ &=&-{1\over(4\pi)^2}
\left({\mu\over2m}\right)^{4\epsilon}
\left[
{3m\over p}\arctan{p\over3m}+{1\over2}\ln{p^2+9m^2\over64m^2}
+{\cal O}(\epsilon)
\right]\;.
\label{sundiff}
\eqa
The first term $I_{\rm ds1}$ is finite in three dimensions.
Using Eq.~(\ref{sundiff}), we obtain
\bqa
I_{\rm ds1}&=&{1\over2m(4\pi)^5}
\left({\mu\over2m}\right)^{10\epsilon}\left[
6\log^2\!2 - 6 \log3 + 4\log2 + 5{\rm Li}_2(\mbox{$1\over4$})
+\cal O(\epsilon)
\label{ds1f}
\right]\;.
\eqa
The second term $I_{\rm ds2}$ can be written as
\bqa
I_{\rm ds2}&=&2I_{\rm sun}(p=im)\int_{pqr}{1\over(p^2+m^2)^2}{1\over q^2+m^2}
{1\over r^2+m^2}
{1\over({\bf p}+{\bf q}+{\bf r})^2+m^2}\;.
\eqa
Using Eqs.~(\ref{isunp}) and~(\ref{iballp}), we obtain
\bqa
I_{\rm ds2}
&=&
{1\over16m(4\pi)^5}
\left({\mu\over2m}\right)^{10\epsilon}
\times\nonumber\\&&\times
\left[{1\over\epsilon^2}
+(8-12\log2){1\over\epsilon}
+52+{5\over4}\pi^2-96\log2
+44\log^22
+\cal O(\epsilon)\right]\;.
\label{ds2f}
\eqa
Similarly, $I_{\rm ds3}$ can be written as
\bqa\nonumber
I_{\rm ds3}&=&-I_{\rm sun}^2(p=im)I_2 \\
&=&-
{1\over32m(4\pi)^5}\left({\mu\over2m}\right)^{10\epsilon}
\times\nonumber\\&&\times
\left[{1\over\epsilon^2}
+(12-16\log2){1\over\epsilon}
+108-{\pi^2\over12}-192\log2+80\log^22
+\cal O(\epsilon)
\right].
\label{ds3f}
\eqa
Adding Eqs.~(\ref{ds1f}),~(\ref{ds2f}), and~(\ref{ds3f}), we obtain
\bqa\nonumber
I_{\rm doublesun}&=&{1\over32m(4\pi)^5}\left({\mu\over2m}\right)^{10\epsilon}
\times\nonumber\\&&\times
\bigg[
{1\over\epsilon^2}
+\left(4-8\log2\right){1\over\epsilon}-4+{31\over12}\pi^2
-96\log3+64\log2+104\log^22
 +\nonumber\\
&&\hphantom{\times\bigg[}+80{\rm Li}_2(\mbox{$1\over4$})
+\cal O(\epsilon)\bigg]\;.
\eqa
Let us finally discuss the five-loop integral appearing in 
Eq.~(\ref{i5c}). It can be written as
\bqa
I_{\rm 4j}&=&\int_{pq}I_{\rm bubble}(p)\left[\Pi_{\rm tri}(p,q)\right]^2\;,
\label{4j}
\eqa
where 
\bqa
\Pi_{\rm tri}(p,q)&=&\int_{r}{1\over r^2+m^2}{1\over({\bf p}+{\bf r})^2+m^2}
{1\over({\bf q}+{\bf r})^2+m^2}
\label{selftri}
\;.
\eqa
The diagram~(\ref{selftri}) is finite in three dimensions and can be written
as~\cite{Nickel:1978ds,kast:2009}
\bqa
\Pi_{\rm tri}(p,q)&=&{\arctan(\sqrt{D}/C)\over8\pi\sqrt{D}}\;,
\eqa
where
\bqa
C&=&{p^2+q^2+{\bf p\cdot q}+4m^2\over m^2}\,,
\\
D&=&{p^2q^2({\bf p}-{\bf q})^2+4m^2[p^2q^2-({\bf p\cdot q})^2]\over4m^6}\;.
\eqa
The integral~(\ref{4j}) can now be evaluated numerically 
by first averaging over angles and then integrating over $p$ and $q$.
This yields
\bqa
I_{\rm 4j}&=&
{1\over m(4\pi)^5}\left({\mu\over2m}\right)^{10\epsilon}
\left[0.443166\right]\;.
\eqa

\bibliography{ntnu}
\bibliographystyle{JHEP}

\end{document}